\documentclass[preprint,amssymb,onecolumn]{revtex4-1}%

\usepackage{graphicx}
\usepackage{epsfig}
\usepackage{amsmath}
\usepackage{graphics}
\usepackage{color}
\usepackage{amsmath,amssymb}

\begin{document}

\title{Strong connection between single-particle and density excitations \\ in Bose--Einstein condensates}

\author{Shohei Watabe}  
\affiliation{
Department of Physics, Faculty of Science Division I, Tokyo University of Science, Shinjuku, Tokyo,162-8601, Japan
}

\begin{abstract} 

Strong connection between the single-particle excitation and the collective excitation stands out as one of the features of Bose--Einstein condensates (BECs). We discuss theoretically these single-particle and density excitations of BECs focusing on the exact properties of the one-body and two-body Green's functions developed by Gavoret and Nozi\`eres. We also investigate these excitations by using the many-body approximation theory at nonzero temperatures. First, we revisited the earlier study presented by Gavoret and Nozi\`eres, involving the subsequent results given by Nepomnyashchii and Nepomnyashchii, in terms of the matrix formalism representation. This matrix formalism is an extension of the Nambu representation for the single-particle Green's function of BECs to discuss the density and current response functions efficiently. We describe the exact low-energy properties of the correlation functions and the vertex functions, and discuss the correspondence of the spectra between the single-particle excitation and the density excitation in the low-energy and low-momentum limits at $T=0$. After deriving the exact low-energy structures of the one-body and two-body Green's functions, we develop a many-body approximation theory of BECs with making the use of the matrix formalism for describing the single-particle Green's function and the density response function at nonzero temperatures. We show how the peaks of the single-particle spectral function and the density response function behave with an increasing temperature. Many-body effect on the single-particle spectral function and the density response function is included within a random phase approximation, where satellite structures emerge because of beyond-mean-field effects. Criticisms are also made on recent theories casting doubt upon the conventional wisdom of the BEC: the equivalence of the dispersion relations between the single-particle excitation and the collective excitation in the low-energy and low-momentum regime. 
 
\end{abstract}

\maketitle

\section{Introduction} 

One of the motives for the study of the condensed matter physics is to know excitations in a quantum many-body system, which provides deep understandings of physics behind the system~\cite{abrikosov1975methods,pines1966theory,nozieres1990theory,mahan2000many}. 
Various kinds of response functions, such as the single-particle spectral function, density response function, pair-correlation function and spin response function, are useful to understand excitations including the single-particle excitation, density excitation, pair-breaking, and spin excitation. 
Generally, even if the two-body correlation function is constructed from the one-body correlation function, the peak structure of the single-particle excitation does not directly clearly emerge as the exact same peak structure in the density response function generated from the two-body correlation function, where effect of the single-particle excitation may emerge as the broad continuum~\cite{abrikosov1975methods,pines1966theory,nozieres1990theory,mahan2000many}. 
In contrast to this wisdom, Bose--Einstein condensates (BECs) is of particular interest, since the single-particle property strongly relates to the collective property~\cite{griffin1993excitations}. 
The Josephson sum-rule concludes that the outcome of the coherent flow of particles is related to the single-particle spectral function, which explicitly gives the relation among the superfluid density, the condensate density, and the single-particle Green's function~\cite{JOSEPHSON1966608,Holzmann:2007baa,ueda2010fundamentals}. 
Gavoret and Nozi\`eres also provided the exact result which states that in the low-energy and low-momentum regime at absolute zero temperature, the density response function shares the pole of the single-particle spectral function, and both the single-particle and collective excitations are the phonon, the speed of which is equal to the thermodynamic compressible sound mode~\cite{Gavoret:1964gv}. 

Excitations in the superfluid helium have been studied extensively and intensively~\cite{nozieres1990theory,griffin1993excitations,pitaevskii2016bose}, including the phonon excitation strongly related to the Landau's criterion for stability of the superfluid, roton and maxon excitations which gives the minimum and maximum in the dispersion law, and hydrodynamic modes such as the first sound and second sound. 
The liquid helium is highly correlated system with a large gas parameter, where the Bogoliubov theory cannot be directly applied to compare the experimental results. 
Ultracold atomic gases with a small gas parameter have been a preferable play ground to test the mean-field theory described by the Gross--Pitaevskii equation~\cite{Gross:1961tq,PitaevskiiLP:1961tf} and the Bogoliubov theory~\cite{Bogolyubov:1947zz}. 
Furthermore, the recent experimental realization of the box trap in ultracold gases~\cite{Gaunt:2013ip} releases us from the conventional restriction of harmonic trap effects, which opens the study of the quantum many-body physics in a highly controllable uniform system.  
Through the significant development of the field of ultracold atomic gases~\cite{Leggett2001,Dalfovo1999,stoof2008ultracold,pitaevskii2016bose,w2017quantum}, 
Feshbach resonance can be used to tune the interaction strength from the weakly interaction to the strongly interaction~\cite{Chin:2010kl}, the phase contrast image can measure the density fluctuation in the real space~\cite{Andrews1997}, and the Bragg spectroscopy can measure the structure factor in the momentum space as well as the energy space~\cite{Stenger1999,Stamper-Kurn1999,Steinhauer:2002hgb}. 
The recent experiments of the BEC in the ultracold atomic gases have expanded the scope beyond the mean-field region~\cite{Papp:2008iu}. 
The ultracold atomic gases may also serve as an ideal potential platform for directly addressing the strong connection between the single-particle excitation and the density excitation in BECs. 
On the other hand, several theories have been proposed that cast doubt on the paradigm about the BEC~\cite{Navez:2008ig,Navez:2010dz,Kita:2010fv,Kita:2011gqb,Kita:2014bc,Kita:2019if,Tsutsumi2016}: the correspondence between the single-particle excitation and the collective excitation in the low-energy and low-momentum region.

The tour de force by Gavoret and Nozi\`eres proves the simple exact property of the BEC at the absolute zero temperature~\cite{Gavoret:1964gv}; 
the density response function shares the pole of the single-particle spectral function, which gives the phonon excitation with the thermodynamic compressible sound speed. 
To follow their proof, we face two separate tasks; 
One is to analyze and to relate diagrammatic structures of the ground state energy, self-energy contributions, correlation functions, and vertex functions. 
The other is to calculate relations of diagrammatic structures obtained in the first task by using identities of Green's functions, 
where a few notations were not given in the modern way in the original paper~\cite{Gavoret:1964gv}. 
In this paper, we first revisit the Gavoret--Nozi\`eres analysis by introducing a systematic formalism for the BECs. 
The Bardeen--Cooper--Schrieffer theory for the superconductivity has been well formulated by using the Nambu representation, which successfully discusses the gauge invariance and the Meissner effect in the theory of superconductivity~\cite{Nambu:1960bz,schrieffer1964theory}. 
Although the BEC theory has been also formulated by using the Nambu representation, 
the theory of density and current correlation functions in the BEC does not fully benefit from the Nambu representation. 
We reconstruct the BEC theory for the density and current correlation functions given by Gavoret and Nozi\`eres by using the matrix formalism with the extension of the Nambu representation to these correlation functions, which can reproduce exact relations efficiently. 

Recent work~\cite{Watabe:2018id} has investigated the multiparticle excitation in ultracold gases by using a many-body approximation, and also studied the energy and momentum dependence of the single-particle excitation as well as the collective excitation at the nonzero temperature.  
This earlier study employed the approximation that satisfies exact relations, where the off-diagonal self-energy as well as the density vertex for the density response function vanish in the low-energy and low-momentum limits~\cite{Nepomnyashchii:1975vs,Nepomnyashchii:1978wb}. 
However, even if the approximation satisfies these exact identities, which are called the Nepomnyashchii--Nepomnyashchii identity and the zero-frequency density vertex identity, it may not exclude the possibility of the approximation dependence of the results, and it does not guarantee that the approximation satisfying these exact identities reproduces qualitative behaviors as well as quantitative properties of the BEC. 
In this paper, in addition to the study of the exact properties, we also address the single-particle spectral function and density response function by using the many-body approximation theory at nonzero temperatures. 
We take the many-body theory different from the earlier paper~\cite{Watabe:2018id} with focusing on the effect of the vertex corrections, and discuss the qualitative properties common in these approximations.

This paper is structured as follows. 
Section~\ref{SecII} introduces the correlation functions of BECs studied in the present paper. 
Section~\ref{SecNewIII} presents the details of the matrix form for the correlation and vertex functions, which can efficiently address structures of diagrams in BECs. 
Section~\ref{Sec:Low-energy} describes the relations between the vertex functions in the low-energy regime. 
Using these results, the low-energy behaviors of the correlation functions at $T=0$ are discussed in Sec.~\ref{SecIV}. 
The formulation in these sections, where the earlier result by Gavoret and Nozi\`eres~\cite{Gavoret:1964gv} and the subsequent result by Nepomnyashchii and Nepomnyashchii~\cite{Nepomnyashchii:1975vs,Nepomnyashchii:1978wb} are revisited, is developed in the matrix formalism. 
This formalism can efficiently discuss structures of diagrams and infrared divergences in BECs~\cite{Watabe:2013hw,Watabe:2014gv,Watabe:2018id,Watabe:2019fz}, which has been successfully applied to reproduce the Nepomnyashchii--Nepomnyashchii identity~\cite{Watabe:2014gv}.
Section~\ref{SecV} reviews the earlier experimental and theoretical studies focused on the sound mode in the superfluid helium as well as ultracold atomic BECs, where variant sound modes in the superfluid, such as the second sound, are important but beyond the scope of the present paper. 
This section also serves as criticisms of recent theories casting doubt upon the paradigm about the BEC: the equivalence of the dispersion relations between the single-particle excitation and the collective excitation in the low-energy and low-momentum regime. 
Section~\ref{SecVI} develops the formulation of the random phase approximation in terms of the matrix formalism. 
Section~\ref{SecVII} discusses the single-particle spectral function and the density response function at nonzero temperatures within the many-body approximation developed in the previous section~\ref{SecVI}. 
This section also addresses the correspondence between peaks of the single-particle spectral function and the density response function, and studies the sound speed estimated from the compressibility zero-frequency sum-rule by using the density response function obtained in the random phase approximation. 
We end with the summary and conclusions in Sec.~\ref{SecVIII}. 

Throughout this paper, we set $\hbar = k_{\rm B} = 1$, and take the system volume $V$ to be unity.   
The terms, $n$-particle irreducible ($n$PI) and $n$-particle reducible ($n$PR), are applied to represent diagrams that cannot and can be separated into two pieces by cutting $n$ single-particle lines, respectively. 
The regular part called in this paper means the proper part, which represents a diagram that cannot be separated into two pieces by cutting a single interaction line.

\section{correlation functions}\label{SecII}
 
We consider the Hamiltonian of an interacting Bose system with the atomic mass $m$, given by 
\begin{align}
\hat {H} = &
\int d {\bf r} 
\frac{1}{2m} \nabla 
\hat \Psi^{\dag} ({\bf r})
\nabla \hat \Psi ({\bf r})
\nonumber 
\\
& 
+ \frac{1}{2} 
\int d {\bf r} \int d {\bf r}' 
\hat \Psi^{\dag} ({\bf r})
\hat \Psi^{\dag} ({\bf r}')
U_{\rm int} ({\bf r} - {\bf r'})  
\hat \Psi^{} ({\bf r}')
\hat \Psi^{} ({\bf r}), 
\label{eq1}
\end{align} 
where $\hat \Psi ({\bf r})$ and $\hat \Psi^{\dag} ({\bf r})$ are bosonic annihilator and creator, respectively. 
In the BEC ordered phase, 
the field operator may be treated by the so-called Bogoliubov prescription: 
\begin{align}
\hat \Psi ({\bf r}) = & \Phi_{0}  ({\bf r}) + \hat \phi ({\bf r}), 
\label{eq2}
\end{align} 
where $\Phi_{0} ({\bf r})$ represents the order parameter of the condensate wave function, and 
$\hat \phi ({\bf r})$ the non-condensate part. 
In the uniform system with the condensate density $n_{0}$, we may suppose 
\begin{align}
\hat \Psi ({\bf r}) = & \sqrt{n_{0}} + \sum\limits_{{\bf p}\neq 0} \hat a_{\bf p} \exp{(i {\bf p} \cdot {\bf r})}, 
\label{eq3}
\end{align}
where the condensate wave function is taken to be real, i.e., $\Phi_{0} ({\bf r}) = \sqrt{n_{0}}$. 

We consider a contact interaction $U_{\rm int} ({\bf r} - {\bf r}') = U \delta ({\bf r} - {\bf r}')$, where the interaction strength $U$ is related to the $s$-wave scattering length $a_{s}$ through the relation 
$4 \pi a_s/m = U/ [1+ \displaystyle{U} \sum\limits_{{\bf p}}^{p_{\rm c}} 1/(2 \varepsilon_{\bf p})  ]$, 
where $p_{\rm c}$ is the cutoff momentum, and $\varepsilon_{\bf p}$ the kinetic energy of the bosonic particle $\varepsilon_{\bf p} = {\bf p}^{2}/2m$. 

An average of an operator $\hat O$ at temperature $T$ is given by 
$\langle \hat O \rangle = {\rm Tr} [ \exp{ (- \hat {\mathcal H}/T )} \hat O ] / \Xi$. 
Here, the Hamiltonian $\hat {\mathcal H}$ with the chemical potential $\mu$ is given by 
$\hat {\mathcal H} = \hat H - \mu \int d {\bf r} \hat \phi^{\dag} ({\bf r}) \hat \phi ({\bf r})$, 
where the Bogoliubov prescription is applied to $\hat H$. 
The partition function is given by $\Xi = {\rm Tr}[\exp{(- \hat {\mathcal H}/T)}]$, 
which may be regarded as the quasi-grand partition function because the term $-\mu n_{0}$ is omitted from the hamiltonian $\hat {\mathcal H}$. It is sufficient to evaluate an average by using $\exp{(- \hat {\mathcal H}/T)}$ with the Bogoliubov prescription, because the term $-\mu n_{0}$ is the $c$-number, which is reduced in the form of the average. 

We introduce three representations of the single-particle thermal Green's function 
\begin{align}
G ({\bf p}, i \omega_{n}) = & - 
\int_{0}^{1/T} d \tau e^{i \omega_{n} \tau}
\langle T_{\tau} \hat{\bf A}_{\bf p} (\tau) \hat{\bf A}_{\bf p}^{\dag} (0) \rangle, 
\label{eq7}
\\ 
{\bf G} ({\bf p}, i \omega_{n}) = & - 
\int_{0}^{1/T} d \tau e^{i \omega_{n} \tau}
\langle T_{\tau} \hat{\bf A}_{\bf p} (\tau) \otimes \hat{\bf A}_{-\bf p}  (0) \rangle, 
\label{eq8}
\\ 
{\bf G}^{\dag} ({\bf p}, i \omega_{n}) = & - 
\int_{0}^{1/T} d \tau e^{i \omega_{n} \tau}
\langle T_{\tau} \hat{\bf A}_{\bf p}^{\dag} (0) \otimes \hat{\bf A}_{-\bf p}^{\dag}  (\tau) \rangle, 
\label{eq9}
\end{align} 
where $\hat{\bf A}_{\bf p} (\tau) \equiv ( \hat{a}_{\bf p} (\tau ), \hat{a}_{- \bf p}^{\dag} (\tau ) )^{\rm T}$
in the Nambu representation~\cite{Hohenberg:1965}. 
Here, $\otimes$ is the Kronecker product, and $T_{\tau}$ denotes an operation of $\tau$-ordering, 
which arranges operators from right to left in order of increasing the imaginary time $\tau$. 
In the bosonic case, the Matsubara frequency is $\omega_{n} = 2 \pi n T$ with $n \in \mathbb{Z}$. 
The Green's functions $G$, ${\bf G}$ and ${\bf G}^{\dag}$ are $(2 \times 2)$, $(4 \times 1)$, and 
$(1 \times 4)$-matrices, respectively. 

The Dyson equations for the Green's functions are given by 
\begin{align}
G (p) = & G_{0} (p) + G_{0} (p) \Sigma (p) G (p), 
\label{eq11}
\\ 
{\bf G} (p) = & {\bf G}_{0} (p) + \{ [G_{0}(p) \Sigma (p)] \otimes \sigma_{0} \} {\bf G} (p), 
\label{eq12}
\\ 
{\bf G}^{\dag} (p) = & {\bf G}_{0}^{\dag} (p) + {\bf G}^{\dag} (p) \{ \sigma_{0} \otimes [\Sigma (-p) G_{0}(-p) ] \}, 
\label{eq13}
\end{align} 
with $p = (i \omega_{n}, {\bf p})$. Each matrix equation provides equivalent equations for the Green's function $G_{ij}$ with $i,j = 1,2$. Here, $G_{0}^{-1} (p) = i \omega_{n} \sigma_{3} - \varepsilon_{\bf p} + \mu$ is the $(2\times 2)$-matrix Green's function for non-interacting bosons, 
where $\sigma_{j=0, 1,2,3}$ are the Pauli matrices. 
The identity matrix of the size $2$ is given by $\sigma_{0}$. 
The non-interacting parts ${\bf G}_{0}$ and ${\bf G}_{0}^{\dag}$ are given by 
\begin{align}
{\bf G}_{0} (p) = 
\begin{pmatrix} 0 \\ G^{(0)} (p) \\ G^{(0)} (-p) \\ 0 \end{pmatrix}, \quad 
{\bf G}_{0}^{\dag} (p) = 
(0, G^{(0)} (p), G^{(0)} (-p), 0 ), 
\label{eq14}
\end{align}
where $G^{(0)} (p) = 1/(i \omega_{n} - \varepsilon_{\bf p} + \mu)$.  
Interaction effects are included into the $(2\times2)$-matrix self-energy $\Sigma$, and we may introduce the $(4\times 1)$-matrix self-energy ${\bf \Sigma}$, 
and the $(1\times 4)$-matrix self-energy ${\bf \Sigma}^{\dag}$. 
Diagrammatic representations of matrix elements of $(G, {\bf G}, {\bf G}^{\dag})$ as well as $(\Sigma, {\bf \Sigma}, {\bf \Sigma}^{\dag})$ are summarized in Sec.~\ref{SecNewIII}. 
Matrix elements $G_{ij}$ are not independent of each other: $G_{11} (p) = G_{22} (-p)$ and $G_{12} (p) = G_{21} (p) = G_{12} (-p) = G_{21} (-p)$, 
where the self-energies $\Sigma_{ij}$ satisfy the same relations~\cite{abrikosov1975methods,Gavoret:1964gv}.

The Green's function $G$ provides the non-condensate density 
\begin{equation}
n' = n - n_0 = - \frac{1}{2} T \sum_{p} {\rm Tr} [ G (p) \exp{(\sigma_{3} i \omega_{n} \delta)} ], 
\label{eq17} 
\end{equation} 
where $n$ is the total particle density. 
In the following, we omit the convergence factor $\exp{(\sigma_{3} i \omega_{n} \delta)}$ for simplicity. 
The formalism at $T=0$ is introduced by applying the analytic continuation ($i \omega_{n} \rightarrow \omega + i \delta$) 
as well as the following replacement $- T\sum\limits_{n} \rightarrow i \int  d \omega / (2\pi)$~\cite{abrikosov1975methods,schrieffer1964theory}.

\begin{figure}
\begin{center}
\includegraphics[width=8cm]{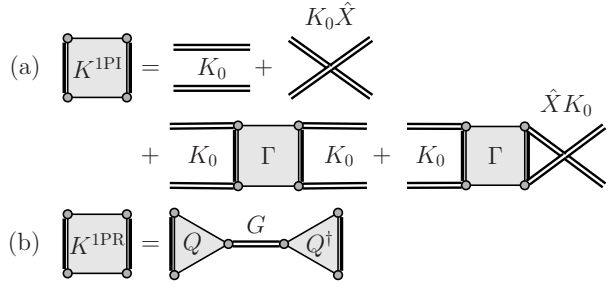} 
\end{center}
\caption{
Two-particle Green's function. 
(a) One-particle irreducible part $K^{\rm 1PI}$. 
(b) One-particle reducible part $K^{\rm 1PR}$. 
}
\label{fig1.fig}
\end{figure}

The $(4\times4)$-matrix two-particle Green's function $K (p, p'; q)$ is composed of the one-particle reducible (1PR) and one-particle irreducible (1PI) parts, i.e., $K^{\rm 1PR}$ and $K^{\rm 1PI}$, where the 1PR part is specific to the condensed Bose system. 
(In Ref.~\cite{Gavoret:1964gv}, these are called the singular and regular parts, respectively.) 
The two-particle Green's function $K$ is given by (See Fig.~\ref{fig1.fig})
\begin{align}
K (p, p'; q) = & K^{\rm 1PI} (p, p'; q) + K^{\rm 1PR} (p, p'; q), 
\label{eq19} 
\end{align}
where 
\begin{align} 
K^{\rm 1PI} (p, p'; q) = & 
K_{0} (p;q) T^{-1}(\delta_{p,p'}  + \hat X \delta_{p,-p'-q} ) 
\nonumber
\\  
& - 
K_{0} (p;q) \Gamma (p, p'; q) K_{0} (p';q) 
\nonumber
\\ 
& - 
K_{0} (p;q) \Gamma (p, -p'-q; q) \hat X K_{0} (p';q), 
\label{eq20} 
\\ 
K^{\rm 1PR} (p, p'; q) = & -  Q (p;q)  G (q) Q^{\dag} (p';q) . 
\label{eq21} 
\end{align} 

\begin{figure}
\begin{center}
\includegraphics[width=8cm]{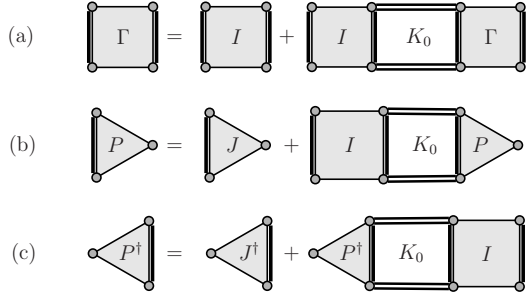} 
\end{center}
\caption{
(a) Bethe--Salpeter equation of the four-point vertex $\Gamma$. Here, $I$ is a two-particle irreducible (2PI) part of the four point vertex. 
(b) Equation of the three-point vertices $P$. (c) Equation of $P^{\dag}$. 
Here, $J$ and $J^{\dag}$ are 2PI parts of the three-point vertices. 
}
\label{fig2.fig}
\end{figure}

For the 1PI part, $K_{0}$ is a bare part of the $(4\times4)$-matrix two-particle Green's function 
\begin{align}
K_{0} (p;q) = G (p+q) \otimes G (-p), 
\label{eq22} 
\end{align}
and $\Gamma$ is the $(4\times4)$-matrix four point vertex, given by (See Fig.~\ref{fig2.fig} (a))
\begin{align}
\Gamma (p,p'; q) = & 
I (p,p'; q) 
- T \sum\limits_{p''}
I (p,p''; q) K_{0} (p'';q) \Gamma (p'', p'; q). 
\label{eq23} 
\end{align}
Here, $I (p,p'; q)$ is a two-particle irreducible (2PI) part of the $(4\times4)$-matrix four point vertex. 
The matrix $\hat X$ is given by 
\begin{align}
\hat X= 
\begin{pmatrix}
1 & 0 & 0 & 0 \\
0 & 0 & 1 & 0 \\
0 & 1 & 0 & 0 \\
0 & 0 & 0 & 1 
\end{pmatrix}, 
\label{eq24} 
\end{align} 
which exchanges upper and lower ends of a two-particle Green's function (See Sec.~\ref{SecNewIII}).

For the 1PR part, the $(4\times2)$- and $(2\times4)$-matrix vertices $Q$ and $Q^{\dag}$ are given by 
\begin{align}
Q (p;q) = & K_{0} (p;q) P (p;q)  
+ T^{-1} \sqrt{-1} ( \delta_{p,0} + \hat X \delta_{p,-q} ) {\mathcal G}_{1/2} ,  
\label{eq25} 
\\
Q^{\dag} (p;q) = & P^{\dag} (p;q) K_{0} (p;q)  
+ T^{-1} \sqrt{-1} {\mathcal G}_{1/2}^{\dag}( \delta_{p,0} + \hat X \delta_{p,-q} ).  
\label{eq26} 
\end{align} 
Here, the $(4\times2)$- and $(2\times4)$-matrix three point vertices $P$ and $P^{\dag}$ are given by (See Figs.~\ref{fig2.fig} (b) and (c))
\begin{align}
P (p;q) 
= & J (p;q) 
- T \sum\limits_{p'} 
I (p, p' ;q) K_{0} (p';q) P (p';q) ,  
\label{eq27} 
\\
P^{\dag} (p;q)  
= & J^{\dag} (p;q) 
- T \sum\limits_{p'} 
P^{\dag} (p';q) K_{0} (p';q) I (p', p;q) ,  
\label{eq28} 
\end{align} 
where $J$ and $J^{\dag}$ are 2PI parts of $P$ and $P^{\dag}$. 
The condensate contributions here are included by the $(4\times2)$ and $(2\times4)$-matrix condensate Green's functions 
\begin{align}
{\mathcal G}_{1/2} = \sigma_{0} \otimes G_{1/2}, \quad 
{\mathcal G}_{1/2}^{\dag} = \sigma_{0} \otimes G_{1/2}^{\dag} , 
\label{eq29} 
\end{align}  
where $G_{1/2}$ and $G_{1/2}^{\dag}$ are the condensate Green's functions 
\begin{align}
G_{1/2} = \sqrt{-1} \begin{pmatrix} \Phi_{0} \\ \Phi_{0}^{*} \end{pmatrix}, 
\quad 
G_{1/2}^{\dag} = \sqrt{-1} ( \Phi_{0}^{*}  , & \Phi_{0}   ). 
\label{eq30} 
\end{align} 
In the case at $T=0$, the factor $\sqrt{-1}$ is replaced with $\sqrt{-i}$~\cite{Hohenberg:1965}.

The density and current correlation functions $\chi_{\mu\nu}^{} (q)$ are defined as 
\begin{align}
\chi_{\mu\nu}^{} (q) 
= & 
- 
\frac{1 }{ 4  }T^{2}
\sum\limits_{p,p'} 
\langle \lambda_{\mu} (p;q) | K (p,p';q) | \lambda_{\nu} (p' ; q)  \rangle, 
\label{eq32} 
\end{align} 
where the density-density and current-current correlation functions are $\chi_{00}^{} (q)$ and $\chi_{ij}^{} (q) $ for $i,j = 1,2,3$, respectively. 
Here, $i,j = 1,2,3$ are the index of the Cartesian coordinate. 
The density-current correlation functions are $\chi_{0i}^{} (q) $ and $\chi_{i0}^{} (q)$ for $i = 1,2,3$. 
The density and current vertex vector $| \lambda_{\mu} (p ; q)  \rangle$ is given by 
\begin{align}
| \lambda_{\mu} (p ; q)  \rangle = \lambda_{\mu} (p ; q) | f_{\mu} \rangle, 
\label{eq33} 
\end{align}
where 
\begin{align}
\lambda_{\mu} (p;q)
= & 
\left \{ 
\begin{array}{ll}
1 &  ( \mu = 0 )
\\[5pt]
{\displaystyle \frac{1}{m} \left ( {\bf p} + \frac{\bf q}{2} \right )_{i} } & (\mu = i = 1,2,3) , 
\end{array} 
\right . 
\label{eq34} 
\end{align} 
and 
\begin{align}
| f_{\mu} \rangle = 
\begin{pmatrix}
0 \\ 1 \\ f_{\mu} \\ 0
\end{pmatrix}, 
\qquad 
f_{\mu} 
= & 
\left \{ 
\begin{array}{ll}
+ 1 &  ( \mu = 0 )
\\[5pt]
-1 & (\mu = i = 1,2,3) . 
\end{array} 
\right . 
\label{eq35} 
\end{align}  
The density vertex vector is simply given by $| \lambda_{0} (p ; q)  \rangle = |f_0 \rangle$. 

\begin{figure}
\begin{center}
\includegraphics[width=8cm]{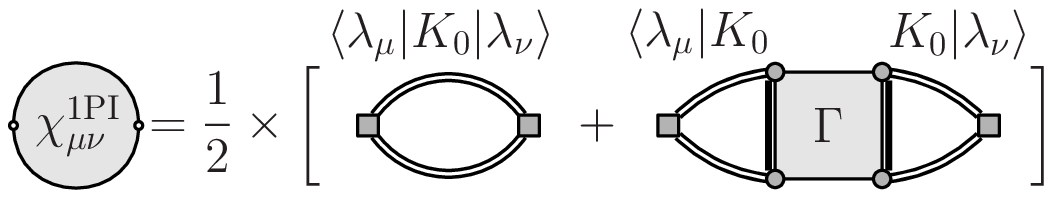} 
\end{center}
\caption{
One-particle irreducible part of the density and current response function $\chi_{\mu\nu}^{\rm 1PI}$. 
}
\label{fig3.fig}
\end{figure}

The correlation functions (\ref{eq32}) are constructed from the two-particle Green's function $K$, which are decomposed into the 1PI and 1PR parts, 
giving the form 
\begin{align}
\chi_{\mu\nu}^{} (q) = \chi_{\mu\nu}^{\rm 1PI} (q) + \chi_{\mu\nu}^{\rm 1PR} (q). 
\label{eq36} 
\end{align} 
The 1PI and 1PR parts are of the form (See Fig.~\ref{fig3.fig})
\begin{align}
\chi_{\mu\nu}^{\rm 1PI} (q) 
= & 
- 
\frac{1 }{ 2  }T
\sum\limits_{p} 
\langle \lambda_{\mu} (p;q) | K_0 (p;q) 
\biggl [ 
| \lambda_{\nu} (p;q) \rangle 
\nonumber 
\\
& 
- T \sum\limits_{p'} \Gamma (p,p';q) K_0 (p';q) | \lambda_\nu (p';q)  
\biggr ] , 
\label{eq37new} 
\\ 
\chi_{\mu\nu}^{\rm 1PR} (q)
= & 
\Upsilon_{\mu}^{\dag} (q)  G (q) \Upsilon_{\nu} (q) , 
\label{eq38} 
\end{align} 
where we used relations $\hat X^{2} = 1$, 
$\hat X | \lambda_{\nu} (-p-q;q)  \rangle = |  \lambda_{\nu} (p;q)  \rangle$ and $\hat X K_{0} (-p-q;q) \hat X =  K_{0} (p;q)$, and introduced the density and current vertices 
\begin{align}
\Upsilon_{\nu} (q) = &  
- \frac{1}{2} T 
\sum\limits_{p'}  Q^{\dag} (p';q) | \lambda_{\nu} (p';q)  \rangle , 
\label{eq40} 
\\
\Upsilon_{\mu}^{\dag} (q) = & - \frac{1}{2} T 
\sum\limits_{p}  \langle \lambda_{\mu} (p;q) | Q (p;q) . 
\label{eq41} 
\end{align}

\section{Diagrammatic representations and matrix forms of correlation and vertex functions}\label{SecNewIII}

\begin{figure}[t]
\begin{center}
\includegraphics[width=6cm]{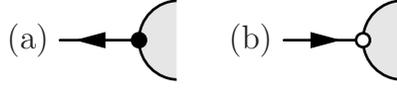}
\end{center}
\caption{Function of vertex points ($\bullet$) and ($\circ$). (a) A filled point ($\bullet$) connects to an outgoing external particle line. 
(b) An open point ($\circ$) connects to an incoming external particle line. A filled point ($\bullet$) can also connect to an open point ($\circ$) and vice versa. 
}
\label{figlist0.fig}
\end{figure}

Correlation and vertex functions are presented in the matrix form in this paper. 
It is very convenient to explicitly provide these representations in terms of diagrams.  
We apply the following rules to satisfy the conservation law. 
The point with the filled circle ($\bullet$) connects to an outgoing external particle line (Fig.~\ref{figlist0.fig}(a)). 
The point with the open circle ($\circ$) connects to an incoming external particle line (Fig.~\ref{figlist0.fig}(b)). 
The point ($\bullet$) can also connect to the point ($\circ$) and vice versa.

\begin{figure}[t]
\begin{center}
\includegraphics[width=8cm]{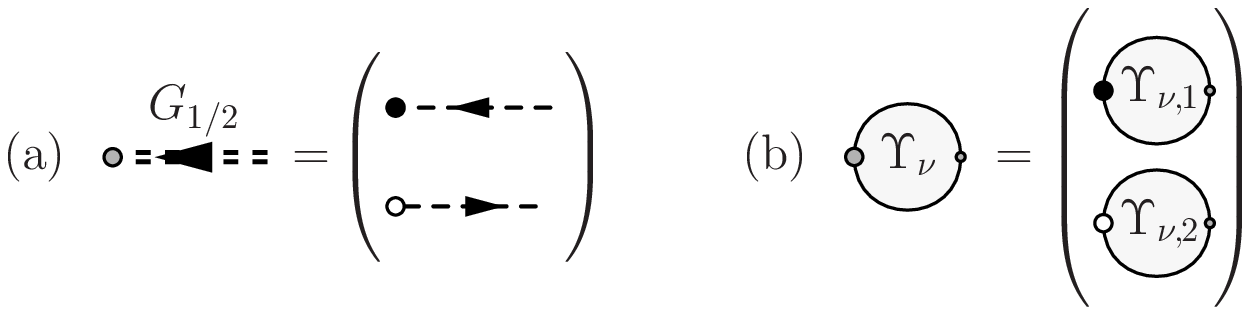} 
\end{center}
\caption{($2\times 1$)-matrix vertex functions. (a) Condensate Green's function $G_{1/2}$. (b) Two point vertex $\Upsilon_\nu $ that connects to an external particle line and an external interaction line. A small grey point connects to an external interaction line.
}
\label{figlist1.fig}
\end{figure} 

\begin{figure}[t]
\begin{center}
\includegraphics[width=9cm]{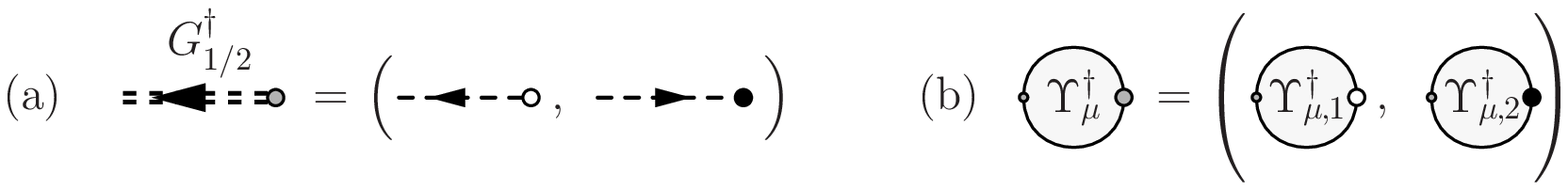} 
\end{center}
\caption{ ($1\times 2$)-matrix vertex functions. (a) Condensate Green's function $G_{1/2}^\dag$. (b) Two point vertex $\Upsilon_\mu^\dag $ that connects to an external particle line and an external interaction line. 
}
\label{figlist2.fig}
\end{figure} 

The ($2\times 1$)-matrix vertex functions include the condensate Green's function $G_{1/2}$, and the two point vertex $\Upsilon_\nu $ that connects to an external particle line and an external interaction line (Fig.~\ref{figlist1.fig}). 
The ($1\times 2$)-matrix vertex functions $G_{1/2}^\dag$ and $\Upsilon_\mu^\dag $ are their counterparts (Fig.~\ref{figlist2.fig}). 

\begin{figure*}[t]
\begin{center}
\includegraphics[width=\textwidth]{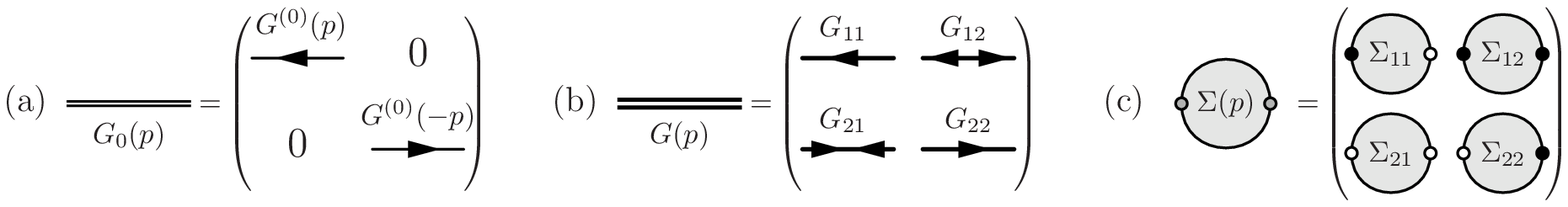} 
\end{center}
\caption{ ($2\times 2$)-matrix correlation and vertex functions. (a) Free-part of the single-particle Green's function $G_0$. (b) Single-particle Green's function $G$. (c) Self-energy $\Sigma$. 
}
\label{figlist3.fig}
\end{figure*}

\begin{figure*}[t]
\begin{center}
\includegraphics[width=\textwidth]{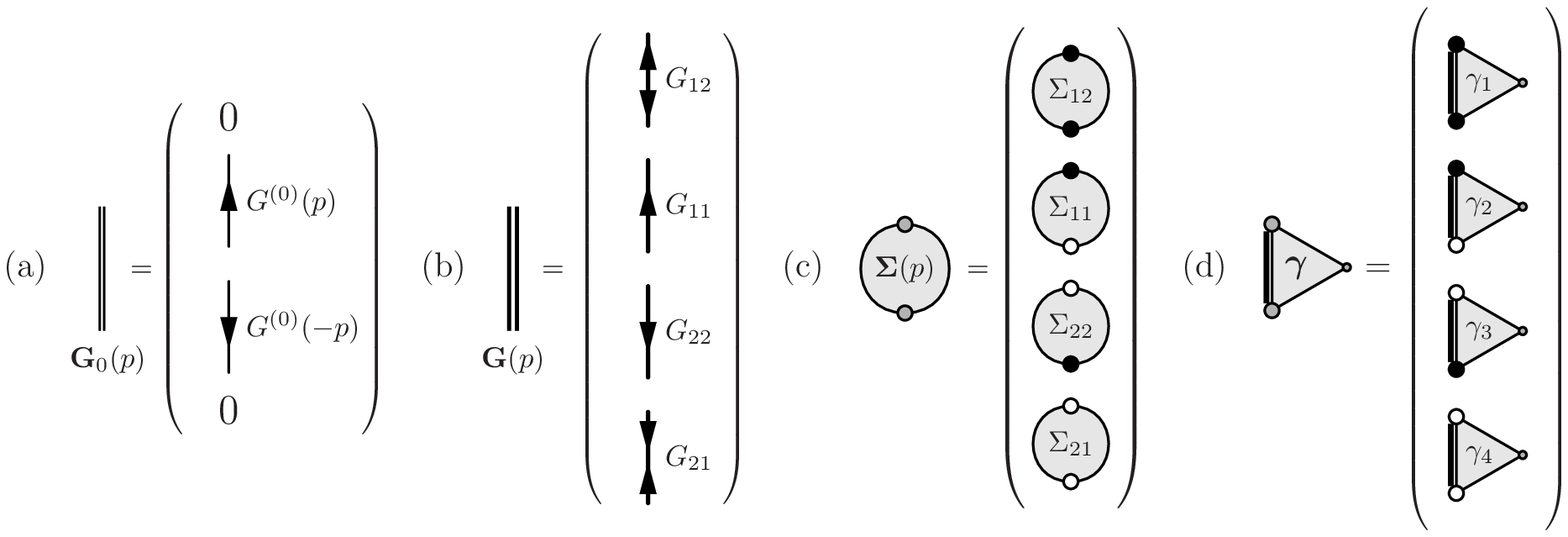}
\end{center}
\caption{ ($4\times 1$)-matrix correlation and vertex functions. (a) Free-part of the single-particle Green's function ${\bf G}_0$. (b) Single-particle Green's function ${\bf G}$. (c) Self-energy ${\bf \Sigma}$. (d) Three-point vertex $\boldsymbol{\mathit \gamma}$ that connects to two external particle lines and an external interaction line. 
}
\label{figlist4.fig}
\end{figure*}

\begin{figure*}[t]
\begin{center}
\includegraphics[width=\textwidth]{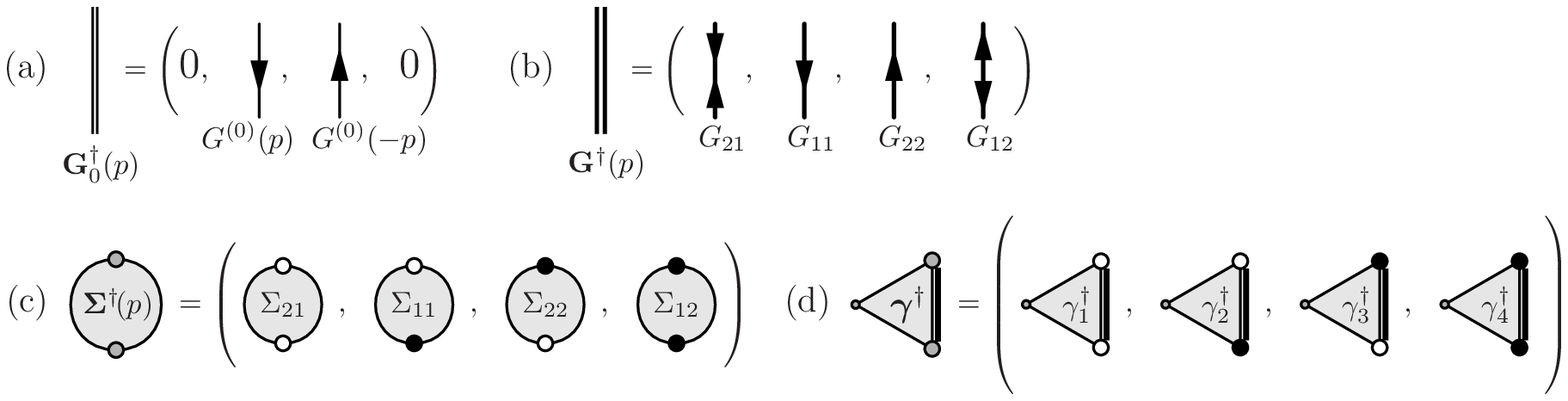}
\end{center}
\caption{ ($1\times 4$)-matrix correlation and vertex functions. (a) Free-part of the single-particle Green's function ${\bf G}_0^\dag$. (b) Single-particle Green's function ${\bf G}^\dag$. (c) Self-energy ${\bf \Sigma}^\dag$. (d) Three-point vertex $\boldsymbol{\mathit \gamma}^\dag$ that connects to two external particle lines and an external interaction line. 
}
\label{figlist5.fig}
\end{figure*}

The ($2\times 2$)-matrix correlation and vertex functions include the single-particle Green's functions $G_0$ and $G$, as well as the self-energy $\Sigma$ (Fig.~\ref{figlist3.fig}). These functions are also given in the ($4\times 1$)- or ($1\times 4$)-matrices. 
The ($4\times 1$)-matrix correlation and vertex functions include the Green's functions ${\bf G}_0$ and ${\bf G}$, the self-energy ${\bf \Sigma}$, as well as the three-point vertex $\boldsymbol{\mathit \gamma}$ that connects to two external particle lines and an external interaction line (Fig.~\ref{figlist4.fig}). The ($1\times 4$)-matrix functions, such as ${\bf G}_0^\dag$, ${\bf G}^{\dag}$, ${\bf \Sigma}^{\dag}$ as well as $\boldsymbol{\mathit \gamma}^\dag$, are their counterparts (Fig.~\ref{figlist5.fig}).

\begin{figure*}[t]
\begin{center}
\includegraphics[width=\textwidth]{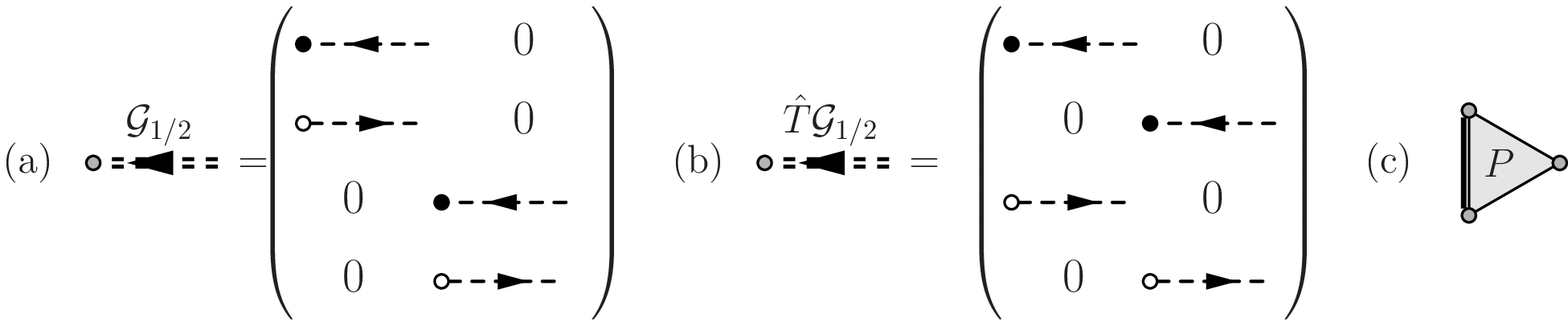}
\end{center}
\caption{ The ($4\times 2$)-matrix vertex functions. (a) Condensate Green's function ${\mathcal G}_{1/2} = \sigma_0 \otimes G_{1/2}$. (b) $\hat X {\mathcal G}_{1/2} = G_{1/2} \otimes \sigma_0$. (c) Three point vertex $P$ that connects to three external particle lines. 
The matrix $Q$ as well as the 2PI part $J$ of the three point vertex $P$ are also given in the same matrix form. 
}
\label{figlist6.fig}
\end{figure*}

\begin{figure*}[t]
\begin{center}
\includegraphics[width=\textwidth]{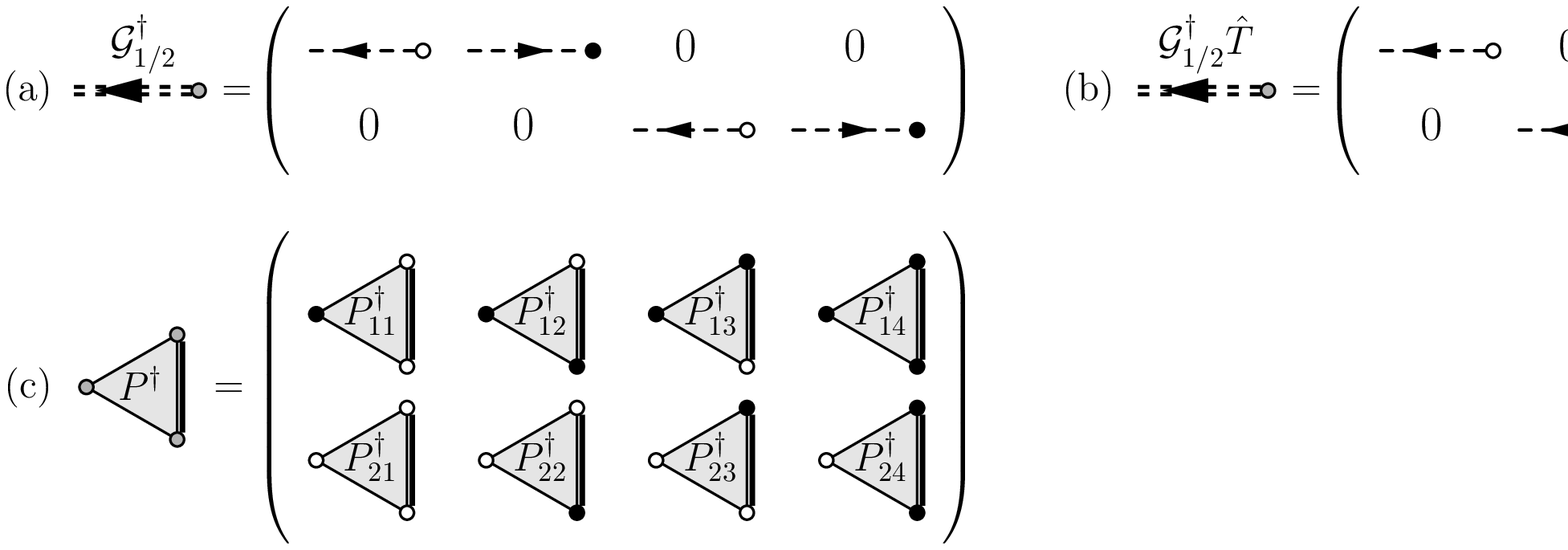}
\end{center}
\caption{
The ($2\times 4$)-matrix vertex functions. (a) Condensate Green's function ${\mathcal G}_{1/2}^\dag = \sigma_0 \otimes G_{1/2}^\dag$. (b) ${\mathcal G}_{1/2}^\dag \hat X = G_{1/2}^\dag \otimes \sigma_0$. (c) Three point vertex $P^\dag$ that connects to three external particle lines. 
The matrix $Q^{\dag}$ as well as the 2PI part $J^{\dag}$ of the three point vertex $P^{\dag}$ are also given in the same matrix form. 
}
\label{figlist7.fig}
\end{figure*}

\begin{figure*}[t]
\begin{center}
\includegraphics[width=16cm,angle=0]{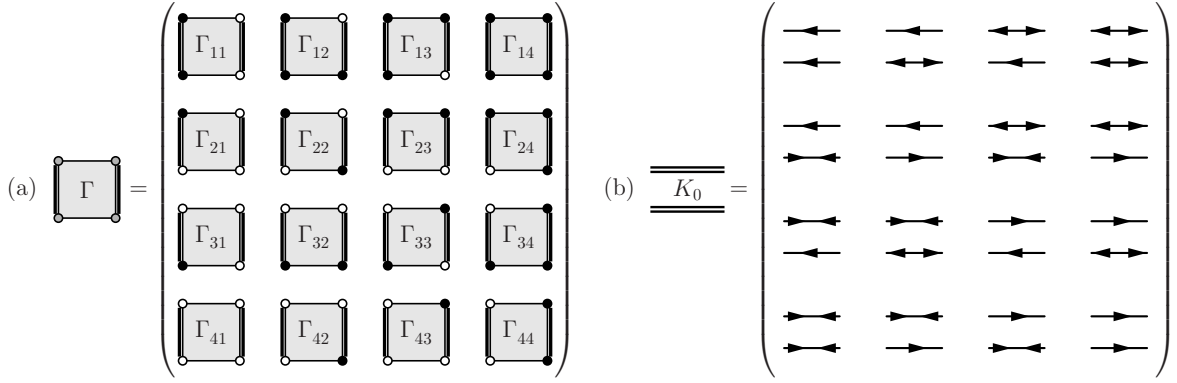}
\end{center}
\caption{ 
($4\times 4$)-matrix correlation and vertex functions. (a) Four-point vertex ${\Gamma}$ that connects to four external particle lines. The 2PI part $I$ of the four point vertex $\Gamma$ is also given in the same matrix form. (b) Bare part of the two-particle Green's function $K_0$.
}
\label{figlist8.fig}
\end{figure*}

\begin{figure}[t]
\begin{center}
\includegraphics[width=8cm,angle=0]{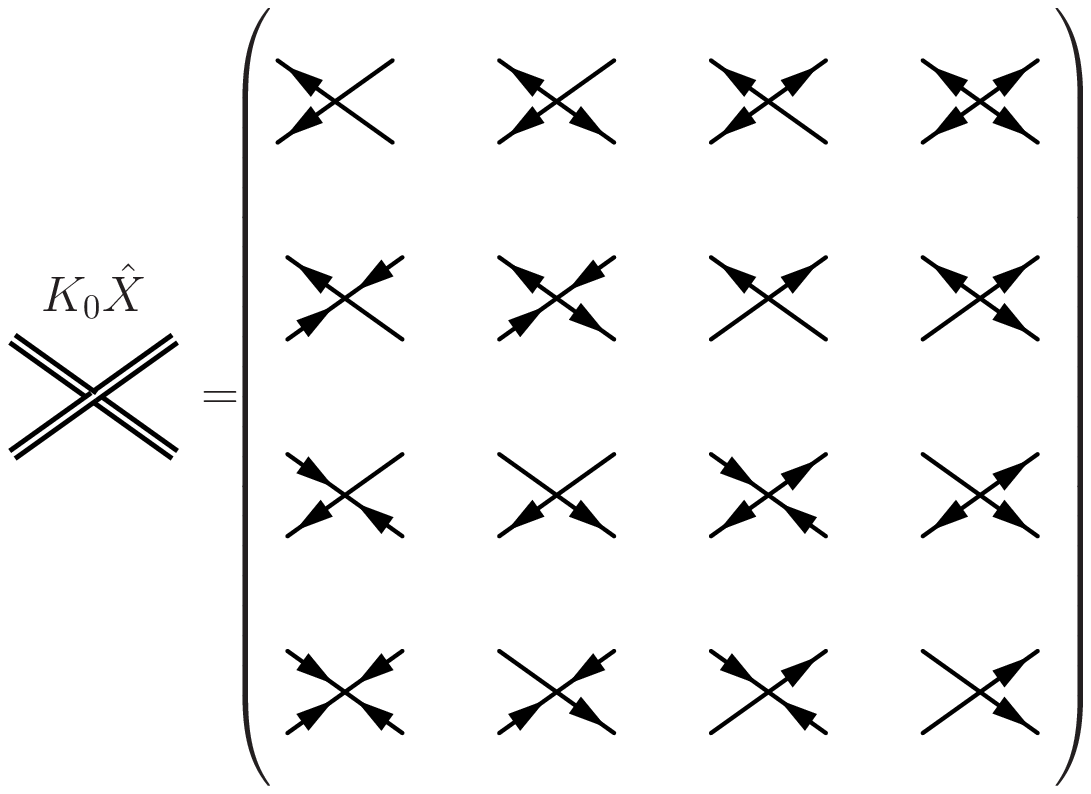}
\end{center}
\caption{ 
Effect of the matrix $\hat X$ on bare part of the two-particle Green's function $K_0$. 
}
\label{figlist8c.fig}
\end{figure}

The ($4\times 2$)-matrix vertex functions include the condensate Green's functions ${\mathcal G}_{1/2} (= \sigma_0 \otimes G_{1/2})$ and $\hat X {\mathcal G}_{1/2} (= G_{1/2} \otimes \sigma_0)$, as well as the three point vertex $P$ that connects three external particle lines (Fig.~\ref{figlist6.fig}). The ($2\times 4$)-matrix vertex functions are their counterparts such as ${\mathcal G}_{1/2}^{\dag} ( = \sigma_0 \otimes G_{1/2}^\dag)$, ${\mathcal G}_{1/2}^{\dag}\hat X (= G_{1/2}^\dag \otimes \sigma_0)$ as well as $P^{\dag}$ (Fig.~\ref{figlist7.fig}). 
The ($4\times 4$)-matrix correlation and vertex functions include the four-point vertex ${\Gamma}$, as well as the two-particle Green's functions, such as $K$, $K_0$, $\hat X K_0 $  and $K_0 \hat X$ (Figs.~\ref{figlist8.fig} and ~\ref{figlist8c.fig}). 

The matrix $\hat X$ provides exchange contributions of the two-particle Green's function. In the diagram of $\hat X K_{0}$ ($K_{0} \hat X$), 
upper and lower left (right) ends of the two-particle Green's function $K_{0}$ are exchanged (Fig.~\ref{figlist8c.fig}). 
The condensate Green's function ${\mathcal G}_{1/2}^{\dag}$ (${\mathcal G}_{1/2}^{} $) connects to lower left (right) corner of the four-point vertex $\Gamma$. 
The condensate Green's function ${\mathcal G}_{1/2}^{\dag}\hat X$ ($\hat X {\mathcal G}_{1/2} $) connects to upper left (right) corner of $\Gamma$. The matrix $\hat X$ also works as $\hat X (\sigma_{0} \otimes G_{1/2}) = G_{1/2} \otimes \sigma_{0}$ and $(\sigma_{0} \otimes G_{1/2}^{\dag}) \hat X = G_{1/2}^{\dag} \otimes \sigma_{0}$. 

The density and current correlation functions $\chi_{\mu\nu}$ are obtained 
by multiplying the two-particle Green's function $K$ by the density/current vertex vectors $\langle \lambda_\mu |$ and $| \lambda_\nu \rangle$ from its left- and right-hand sides, respectively. 
In the diagrammatic representation, $\langle \lambda_\mu |$ (or $| \lambda_\nu \rangle$) closes leftmost (or rightmost) of $K$ with multiplying it by the vertex function $\lambda_\mu$ (or $\lambda_\nu$) (Fig.~\ref{fig3.fig}). 
In particular, since $| \lambda_0 \rangle = | f_0 \rangle$, the density response function are obtained by multiplying $K$ by $\langle f_0 |$ and $| f_0 \rangle$ from its left- and right-hand sides, respectively, which generate vertex points connecting to the external interaction line.  
These factors $\langle f_0 |$ and $| f_0 \rangle$ also provides the relations 
${\mathcal G}_{1/2}^\dag| f_0 \rangle = {\mathcal G}_{1/2}^\dag \hat X | f_0 \rangle  = G_{1/2}$ and 
$ \langle f_0 | {\mathcal G}_{1/2}  =  \langle f_0 | \hat X {\mathcal G}_{1/2} = G_{1/2}^\dag$, 
which will be useful for calculating density vertices.

\section{Relations between vertex functions}\label{Sec:Low-energy}

Vertex functions in the static and zero-momentum limits can be systematically generated from all the possible linked diagrams that construct the thermodynamic potential $\Omega' = - T \ln \Xi$. 
An exact many-line vertex $M ( n_{\rm out}, n_{\rm in}, n_{U})$ is given by~\cite{Nepomnyashchii:1978wb}  
\begin{align} 
&M ( n_{\rm out}, n_{\rm in}, n_{U} ) =  n_{0}^{(n_{\rm out} - n_{\rm in}) /2 }
\nonumber
\\
& \times \left ( - \frac{\partial }{\partial \mu} \right )_{T,n_{0}}^{n_{U}} \left (  \frac{\partial }{\partial n_{0}} \right )_{T,\mu}^{n_{\rm out}}  
\left [ n_{0}^{n_{\rm in}} \left (  \frac{\partial }{\partial n_{0}} \right )_{T,\mu}^{n_{\rm in}} \right ] \Omega' (T, \mu, n_{0}) . 
\label{eq44}
\end{align}  
Here, $n_{\rm in (out)}$ is the number of incoming (outgoing) particle lines that can connect to the vertex function $M$, and $n_{U}$ is the number of external interaction lines $U$ that can also connect to the vertex function $M$. 
The operator $n_0^{n_{\rm in(out)}/2} (\partial / \partial n_0 )^{n_{\rm in(out)}}$ 
works as the elimination of the $n_{\rm in(out)}$ condensate lines $\Phi_0^{(*)} ( = \sqrt{n_0} )$ from the linked diagrams. 
The operator $(- \partial/\partial \mu)^{n_U}$ affects on the Green's function $G_0$ in the linked diagrams, which provides the $n_U$ vertex points for the interaction line due to the relation $- \partial G_0 / \partial \mu = G_0^2$. 
This prescription was originally invented for the ground state energy at $T=0$~\cite{Nepomnyashchii:1978wb}. 
Since the liked diagrammatic structures for the thermodynamic potential at nonzero temperature are formally the same as those of the ground state energy~\cite{negele1995quantum}, this prescription is also applied to the nonzero temperature case~\cite{Griffin1981}.

The equation \eqref{eq44} generates the self-energy matrix $\Sigma$, the three point vertex matrix $P$, and the density vertex $\Upsilon_{0}$ in the zero-energy and zero-momentum limits, respectively given by 
\begin{align}
\Sigma (0) = & 
\frac{\partial \Omega'}{\partial n_{0}} \begin{pmatrix} 1 & 0 \\ 0 & 1 \end{pmatrix} 
+ n_{0} \frac{\partial^{2} \Omega'}{\partial n_{0}^{2}} \begin{pmatrix} 1 & 1 \\ 1 & 1 \end{pmatrix} , 
\label{eq45} 
\\[2pt]
P (0;0) = & 
2 \sqrt{n_{0}} \frac{\partial^{2} \Omega'}{\partial n_{0}^{2}} 
\eta
+ 
n_{0}^{3/2} \frac{\partial^{3} \Omega'}{\partial n_{0}^{3}} 
\eta_{\rm a}, 
\label{eq46} 
\\
\Upsilon_{0} (0) = & \sqrt{n_{0}} 
\left ( 1 - \frac{\partial^{2} \Omega'}{\partial\mu \partial n_{0}} \right )
| + \rangle ,  
\label{eq47} 
\end{align} 
where $|\pm \rangle = (1,\pm 1)^{\rm T}$, $\langle \pm | = (1,\pm1)$ and 
\begin{align} 
\eta = 
\begin{pmatrix}
1 & 0 \\ 1 & 1 \\ 1 & 1 \\ 0 & 1
\end{pmatrix} , 
\quad 
\eta_{\rm a} = 
\begin{pmatrix}
1 & 1 \\ 1 & 1 \\ 1 & 1 \\ 1 & 1
\end{pmatrix} . 
\label{eq48}
\end{align} 
Vertices ${\bf \Sigma}$ and $P$ are related with each other, giving the form 
\begin{align}
P (0;0)  |- \rangle = & 
\frac{2}{\sqrt{n_{0}}} \Sigma_{12} (0) 
\eta | - \rangle 
= \frac{2}{\sqrt{n_{0}}} \hat A {\bf \Sigma} (0) , 
\label{eq49}
\end{align} 
where $\hat A = {\rm diag} (1, 0, 0, -1)$.

The thermodynamic potential $\Omega'$ is related to the grand potential $\Omega$ by introducing the chemical potential of the condensate $\mu_0$. 
We have the relation $\Omega = \Omega' - \mu_0 n_0$ with the condition $\mu_{0} = \mu$~\cite{Nepomnyashchii:1978wb}. 
Since the condensate density is determined from the condition $\partial \Omega / \partial n_0 = 0$, 
we have 
\begin{align}
\mu_{0} (T, \mu, n_{0})= \mu = \frac{\partial \Omega'}{\partial n_0}. 
\label{eq51}
\end{align}
Given this relation, we may derive the Hugenholtz-Pines relation $\Sigma_{11} (0) -\Sigma_{12} (0) = \mu$~\cite{Hugenholtz:1959jb} (or its matrix form $\Sigma (0) |- \rangle = \mu |- \rangle$). 
The Nepomnyashchii--Nepomnyashchii identity~\cite{Nepomnyashchii:1978wb,Nepomnyashchii:1975vs}, giving the form 
\begin{align}
\Sigma_{12} (0) = & 
n_{0} \frac{\partial^{2} \Omega'}{\partial n_{0}^{2}} 
=n_{0} \left . \frac{\partial \mu_{0}}{\partial n_{0}} \right |_{T, \mu} = 0, 
\label{eq52}
\end{align} 
reduces the Hugenholtz-Pines relation to the following form 
\begin{align}
\Sigma (0) = \mu. 
\label{eq53}
\end{align} 
The derivation of the Nepomnyashchii--Nepomnyashchii identity with the use of the matrix formalism is summarized in Ref.~\cite{Watabe:2014gv}, 
and physics of this identity can be found in Refs.~\cite{Nepomnyashchii:1975vs,Nepomnyashchii:1978wb,Popov:1979vk,Nepomnyashchii:1983uq,Weichman:1988eb,Giorgini:1992bt,griffin1993excitations,popov2001functional,Dupuis:2011gf,Stoof:2013dva,Watabe:2014gv,Watabe:2019fz}. 
The Nepomnyashchii-Nepomnyashchii identity $\Sigma_{12} (0) = 0$, which is strongly related to the weak infrared divergence of the longitudinal susceptibility caused by the convolution of the phase-phase correlation function~\cite{Popov:1979vk,Watabe:2014gv}, can be obtained from the relation between vertex functions and the nature of the infrared divergence in the self-energy diagrams~\cite{Nepomnyashchii:1975vs,Nepomnyashchii:1978wb,Watabe:2014gv}. 
As a result, this identity is also valid at nonzero temperature~\cite{Griffin1981,Watabe:2014gv}. 
The identity (\ref{eq52}) also provides a relation $P (0;0)  |- \rangle = 0$. 

The Nepomnyashchii--Nepomnyashchii identity also provides the zero-frequency density vertex identity, i.e., the vanishing density vertex in the limit $p = 0$, giving the form 
\begin{align}
\Upsilon_0 (0) = 0. 
\label{eq54}
\end{align} 
This is valid in the isothermal condition, and the derivation is summarized in Appendix~\ref{AppendixB}.

\begin{figure}[tb]
\begin{center}
\includegraphics[width=8cm]{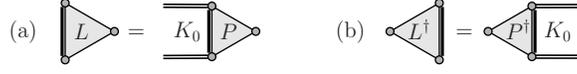} 
\end{center}
\caption{
Three point vertices $L$ and $L^{\dag}$. 
}
\label{fig4.fig}
\end{figure}

In the remaining part of this section, we summarize low-energy behaviors of vertex functions in our matrix representation. 
We first consider a relation between ${\bf \Sigma}$ and $P$ as well as a relation between ${\bf G}$ and $L$, where the ($4 \times 2$)-matrix $L$ (diagrammatically described in Fig.~\ref{fig4.fig}) is given by
\begin{align}
L (p;q) = K_{0} (p;q) P (p;q). 
\label{eq55}
\end{align} 

With respect to $({\bf \Sigma}, P)$ or $({\bf G}, L)$, relations at small but finite $q = ( \varpi, {\bf q})$ are given by~\cite{Gavoret:1964gv} 
\begin{align}
P (p, + q) | 0 \rangle 
+ 
P (p, - q) | 1 \rangle 
\simeq & 
\hat {\mathcal D} (q)
{\bf \Sigma} (p) , 
\label{eq56} 
\\ 
L (p, + q)  | 0 \rangle + L (p, - q)  | 1 \rangle 
\simeq & 
\hat {\mathcal D} (q)
{\bf G} (p) , 
\label{eq57}
\end{align}
where $| 0 \rangle = (1,0)^{\rm T}$, $| 1 \rangle = (0, 1)^{\rm T}$ and $\hat {\mathcal D} (q) =  {\mathcal D}_1 (q) + \hat B {\mathcal D}_2 (q) $
with $\hat B = {\rm diag} (0,-1, +1, 0)$ and 
\begin{align}
{\mathcal D}_1 (q) \equiv & 
\frac{1}{\sqrt{n_{0}}}
\left ( 
2 n_{0} 
 \frac{\partial}{\partial n_{0}}  
+ 
\sum\limits_{\nu= 0}^{3} 
q_{\nu}  \frac{\delta }{\delta x_{\nu}} 
\right ) , 
\\
 {\mathcal D}_2 (q) \equiv & 
\frac{1}{\sqrt{n_{0}}}
\sum\limits_{\nu= 0}^{3} 
q_{\nu}  \partial_{\nu} .
\label{eq59}
\end{align} 
Here, two types of derivatives are introduced: 
$\delta / \delta x_\nu = (\partial/ \partial \mu, \delta / \delta {\bf p})$ 
and 
$\partial_\nu = (\partial / \partial \omega, \partial / \partial {\bf p})$. 
The partial derivative $\partial / \partial p_i$ and the total derivative $\delta / \delta p_{i}$ for $i = 1,2,3$ are respectively defined as 
$\partial G^{(0)} (\pm {\bf p}) / \partial p_i \equiv 
\lim\limits_{d p_i \rightarrow 0} [ G^{(0)} (\pm ( {\bf p} + d p_i {\bf e}_{i}) ) - G^{(0)} (\pm {\bf p} ) ]/ d p_i$, 
and 
$\delta G^{(0)} (\pm {\bf p}) / \delta p_i \equiv 
\lim\limits_{\delta p_i \rightarrow 0} [ G^{(0)} (\pm {\bf p} + \delta p_i {\bf e}_{i} ) - G^{(0)} (\pm {\bf p} ) ]/ \delta p_i$, 
where ${\bf e}_{i}$ is the unit vector in the Cartesian coordinate~\cite{Gavoret:1964gv}. 
The total derivative $\delta / \delta p_{i}$ is related to an observation of the system from a reference frame with a speed $- \delta {\bf p}/m$. 
(Details can be found in Sec. V in Ref.~\cite{Gavoret:1964gv}.)

In the limit $q = 0$, (\ref{eq56}) and (\ref{eq57}) are reduced into 
\begin{align}
\begin{pmatrix}
P (p;0) | + \rangle 
\\
L (p;0) | + \rangle
\end{pmatrix}
= 
2 \sqrt{n_{0}} \frac{\partial}{\partial n_{0}} 
\begin{pmatrix}
{\bf \Sigma} (p)
\\
{\bf G} (p)
\end{pmatrix}. 
\label{eq62}
\end{align}
The three point vertex $P$ (or $L$) is created from the two-point vertex ${\bf \Sigma}$ (or the Green's function ${\bf G}$) 
by eliminating a condensate line from ${\bf \Sigma}$ (or ${\bf G}$) that provides an extra vertex point at $q=0$~\cite{Gavoret:1964gv}. 
 
We also have relations~\cite{Gavoret:1964gv} 
\begin{align}
\begin{pmatrix}
P (p;0) | - \rangle 
\\ 
L (p;0) | - \rangle 
\end{pmatrix}
= 
\frac{2 }{\sqrt{n_{0}}} 
\begin{pmatrix}
\hat A {\bf \Sigma} (p)
\\ 
\hat A {\bf G} (p)
\end{pmatrix} . 
\label{eq63}
\end{align} 
The upper equality in (\ref{eq63}) can be derived as follows~\cite{Gavoret:1964gv}; 
The self-energy ${\bf \Sigma}$ can be constructed from two parts.  
One is the three point vertex $J$, where one of the three vertex points is blocked by a condensate line $\sqrt{n_{0}}$. 
The other is the four point vertex $I$, where two of the four vertex points are blocked by a Green's function $G_{12}$ or $G_{21}$. 
It gives the following relation (See Appendix B in Ref~\cite{Gavoret:1964gv}): 
\begin{align}
2 \hat A {\bf \Sigma} (p) 
= & 
\sqrt{n_{0}} J (p;0) | - \rangle 
 - 
2 T \sum\limits_{p'} I (p,p';0) \hat A {\bf G} (p') . 
\label{eq64}
\end{align} 
By comparing Eqs. (\ref{eq27}) with (\ref{eq64}) with the use of a mathematical identity 
$\hat A  {\bf G} (p) = { K}_{0}  (p; 0) \hat A {\bf \Sigma} (p)$, 
we obtain the first equality of (\ref{eq63}), which is consistent with (\ref{eq49}) at $p=0$. 
The second equality in (\ref{eq63}) with respect to $(L,{\bf G})$ is also obtained by following the similar way. 

According to symmetries, the density and current vertices can be given by  
\begin{align}
\Upsilon_{\nu}^{} (q) = 
\begin{pmatrix}
 \gamma_{\nu} (q) \\ \gamma_{\nu} (-q)
\end{pmatrix}, 
\quad 
\Upsilon_{\mu}^{\dag} (q) 
= 
( \gamma_{\mu} (q) , \gamma_{\mu} (-q)), 
\label{Eq91}
\end{align} 
the matrix element of which in the low-energy regime behaves as (See Appendix~\ref{AppendixC}) 
\begin{widetext} 
\begin{align}
\gamma_{\nu} (q) 
\simeq & 
\sqrt{n_{0}} \lambda_{\nu} (0;q) 
- 
\frac{1}{4} T \sum\limits_{p} 
\lambda_{\nu} (p,0) 
\left \{ 
{\mathcal D}_1 (q)
{\rm Tr} [\sigma_{\nu}' G (p) ] 
-  
{\mathcal D}_2 (q)
{\rm Tr} [\sigma_{\nu}' \sigma_{3} G (p) ] 
\right \}  . 
\label{eq92}
\end{align} 
\end{widetext}  

\section{low energy behaviors of correlation functions at $T=0$}\label{SecIV}
  
The self-energy at small $p$ behaves as~\cite{Nepomnyashchii:1978wb} 
\begin{align}
\Sigma (p) = & \mu + \omega \sigma_{3} + \Delta \Sigma (p) \sigma_+
\nonumber \\ & 
+ \frac{1}{2} \partial_{\omega}^{2} \Sigma ' (0) \omega^{2} + \frac{1}{2} \partial_{{\bf p}}^{2} \Sigma ' (0) {\bf p}^{2} + \cdots , 
\label{eq66}
\end{align} 
where $\sigma_+ = | + \rangle \langle + |$. 
The first term $\mu$ is to satisfy the Hugenholtz--Pines relation as well as the Nepomnyashchii--Nepomnyashchii identity in (\ref{eq53}). 
The term $\Delta \Sigma (p)$ is the so-called non-analytic term~\cite{Nepomnyashchii:1978wb}, which satisfies $\Delta \Sigma (p) \gg |{\bf p}|^{2}$ as well as $\Delta \Sigma (p) \gg \omega^{2}$ at small $p$. 
Note that $\Delta \Sigma (0) = 0$ as well as $\Delta \Sigma (-p) = \Delta \Sigma (p)$ hold, 
because of the Nepomnyashchii--Nepomnyashchii identity $\Sigma_{12} (0) = 0$ and a symmetry relation $\Sigma_{12,21} (-p) = \Sigma_{12,21} (p)$. 
The symmetry relation $\Sigma (0,{\bf p}) = \Sigma (0,-{\bf p}) $ also provides the relation  $\partial_{\bf p} \Sigma ' (0) = 0$, which provides the absence of the first order of ${\bf p}$ in Eq.~\eqref{eq66}. 
Here, the self-energy $\Sigma' (p)$ is defined as $\Sigma ' (p) \equiv \Sigma (p) - \Delta \Sigma (p)\sigma_+$, where the non-analytic term is  extracted.

With respect to the $\omega$-dependence, we have $\partial_\omega \Sigma' (0) = \sigma_3$. 
Because of a symmetry relation $\Sigma_{12} (p) = \Sigma_{12} (-p)$, we find that the off-diagonal element satisfies $\partial_{\omega} \Sigma_{12} ' (0) = 0$. 
For $\partial_\omega \Sigma_{11}' (0)$, we have an identity~\cite{Gavoret:1964gv,Huang:1964bc} 
\begin{align} 
\partial_{\omega} \Sigma_{11}' (0) 
=  \frac{\partial^{2}  \Omega'}{ \partial \mu \partial n_{0}} = 1. 
\label{eq67}
\end{align} 
In the last equality, we have employed the relation (\ref{eqB4}) in the isothermal condition shown in Appendix~\ref{AppendixB}. 
The first equality indicates that the differential $\partial_\omega$ is related to $\partial_\mu$, since the self-energy is constructed from the non-interacting Green's function $G^{(0)}(p) = 1/ (\omega - \varepsilon_{\bf p} + \mu)$ and then the infinitesimally small increase of the energy $\omega + \delta \omega$ in the self-energy $\Sigma$ can be regarded as the infinitesimally small increase of the chemical potential $\mu + \delta \omega$ in a Green's function $G^{(0)}$ that constructs $\Sigma$~\cite{Gavoret:1964gv}. 
With respect to the second order of ${\bf p}$ or $\omega$, similar relations are obtained, giving the forms~\cite{Gavoret:1964gv}  
\begin{align} 
\partial_{\omega}^{2} { \Sigma} ' (0) | - \rangle 
= & 
\frac{1}{n_{0}} \frac{\partial^{2} \Omega'}{\partial \mu^{2}} | - \rangle 
= - \frac{n}{n_0 m c_{\rm T}^2} | - \rangle, 
\label{eq68}
\\ 
\partial_{{\bf p}}^{2} { \Sigma}' (0)  | - \rangle  
= & \frac{n'}{mn_{0}} | - \rangle, 
\label{eq69}
\end{align} 
where $c_{\rm T}$ is the isothermal sound speed (See Appendix~\ref{AppendixB}). 
Note that although the relations between the vertex functions as shown in Sec.~\ref{Sec:Low-energy} hold at nonzero temperatures because the diagrammatic structure is the same as in the case at $T=0$, the relations between the thermodynamic quantities and the differentiations of vertex functions with respect to $p$ shown here may not hold at nonzero temperatures.

The single-particle Green's function in the low-energy regime is reduced into~\cite{Nepomnyashchii:1978wb} (See Appendix~\ref{AppendixBC})
\begin{align} 
G (p) \simeq & \frac{n_{0} m c_{\rm T}^{2}}{n} \frac{ 1 }{ \omega_{}^{2} - c_{\rm T}^{2} {\bf p}^{2} } \begin{pmatrix} 1 & -1 \\ -1 & 1 \end{pmatrix} - \frac{1}{4 \Sigma_{12} (p) } \begin{pmatrix} 1 & 1 \\ 1 & 1 \end{pmatrix}. 
\label{eq70} 
\end{align} 
The first term in (\ref{eq70}) is the leading term of $G$, which provides the phonon spectrum of the single-particle excitation, 
whose sound speed $c_{\rm T}$ corresponds to the isothermal sound speed related to the compressibility (\ref{eqB5}). 
This first term is important to the low-energy behavior of the density and current correlation functions, and essential for the transverse susceptibility $G_{\perp} (p) = - \langle - | G (p) | - \rangle /4 $ with respect to the BEC order parameter~\cite{Giorgini:1992bt,Watabe:2014gv}. 
The second term in (\ref{eq70}) also provides the infrared divergence, because of the Nepomnyashchii-Nepomnyashchii identity, where the infrared divergence of the second term is much weaker than that of the first term in (\ref{eq70}). 
This weak infrared divergent second term never plays an essential role in the density and current correlation functions. 
However, it plays an important role in the longitudinal susceptibility $G_{\parallel} (p) = - \langle +| G (p) | +\rangle /4 $~\cite{Nepomnyashchii:1983uq,Dupuis:2011gf,Dupuis2011,Weichman:1988eb,Giorgini:1992bt,Watabe:2014gv}. 
The transverse and longitudinal fluctuation operators are not commutable~\cite{Weichman:1988eb, Giorgini:1992bt}. 
Since the transverse fluctuation is regarded as the phase fluctuation, the longitudinal fluctuation might be expected to represent the amplitude (Higgs) mode. 
However, the longitudinal susceptibility does not describe the gapped amplitude mode, and shows the weak infrared divergence in the low-energy limit, because it is provided from the convolution of the phase-phase correlation function. 
The response function that can capture the Higgs mode is the scalar susceptibility~\cite{Podolsky:2011ds}.

In the low-energy regime, the 1PI part $\chi_{\mu\nu}^{\rm 1PI} (q)$ can be given by~\cite{Gavoret:1964gv} (See Appendix~\ref{AppendixC}) 
\begin{align}
\chi_{00}^{\rm 1PI} (0) = & \frac{1}{2} T \frac{\partial}{\partial \mu} \sum\limits_{p} {\rm Tr}[G(p)] = - \frac{\partial n'}{\partial \mu} , 
\label{eq84}
\\
\chi_{i 0}^{\rm 1PI} (0) = & \frac{1}{2} T \sum\limits_{p} \frac{p_{i}}{m} \frac{\partial }{\partial \mu}  {\rm Tr}[ \sigma_{3} G(p)] = 0 , 
\label{eq85}
\\
\chi_{0 j}^{\rm 1PI} (0) = & - \frac{1}{2} T \sum\limits_{p} \frac{\delta  }{\delta q_{j}} {\rm Tr}[G(p)] = 0 , 
\label{eq86}
\\
\chi_{ij}^{\rm 1PI} (0) = & - \frac{1}{2} T \sum\limits_{p} \frac{p_{i}}{m} \frac{\delta }{\delta q_{j}} {\rm Tr}[ \sigma_{3} G(p)] = 0 .
\label{eq87}
\end{align}  
The density-density correlation function remains nonzero in the low-energy and low-momentum limit. On the other hand, the density-current and current-current correlation functions vanish in the same limit.

Using \eqref{eq92}, we have the following simple expression of the density and current vertex in the low-energy and low-momentum limits: 
\begin{align} 
\Upsilon_{\nu} (p) \simeq 
\left \{ 
\begin{array}{ll}
{\displaystyle \frac{\omega}{2\sqrt{n_{0}}} \frac{\partial n'}{\partial \mu} | - \rangle } & \quad ( \nu = 0 )
\\[13pt]
{\displaystyle\frac{p_{i}}{2 \sqrt{n_{0}}} \frac{n}{m } | - \rangle } & \quad (\nu = i = 1,2,3) . 
\end{array} 
\right . 
\label{eq93}
\end{align} 
The density and current vertex vanishes in the static and low-momentum limits, i.e., $\Upsilon_{\nu} (0) = 0$, 
which is consistent with the exact identity (\ref{eq54}). 
To obtain (\ref{eq93}), we employed the identity (\ref{eqB4}) in the isothermal condition for $\nu = 0$, which is the consequence of the Nepomnyashchii--Nepomnyashchii identity, and employed the relation $\sum_{\bf p} p_{i} \partial n_{\bf p} / \partial p_{j} = - \delta_{ij} n' $ for  $\nu = i = 1,2,3$, where $n_{\bf p}$ is the Bose distribution function. 

The 1PR parts of correlation functions are then of the forms 
\begin{align}
\chi_{00}^{\rm 1PR} (p) 
= & 
\frac{1}{n_{0}} 
\left ( \frac{\omega}{2} \frac{\partial n'}{\partial \mu}  \right )^{2}
\langle - | G (p) | - \rangle  , 
\label{eq94}
\\ 
\chi_{0i}^{\rm 1PR} (p) 
=
\chi_{i0}^{\rm 1PR} (p) 
= & 
\frac{n \omega p_{i}}{4 m n_{0}} 
\frac{\partial n'}{\partial \mu} 
\langle - | G (p) | - \rangle , 
\label{eq95}
\\
\chi_{ij}^{\rm 1PR} (p) = & 
\frac{n^2}{n_{0}} 
\frac{p_{i}}{2m} 
\frac{p_{j}}{2m} 
\langle - | G (p) | - \rangle , 
\label{eq96}
\end{align} 
which provide  
\begin{align}
\chi_{00}^{\rm 1PR} (p) \simeq & \frac{n}{m} \frac{{\bf p}^{2}}{ \omega^{2} - c_{\rm T}^{2} {\bf p}^{2} } + \frac{\partial n'}{\partial \mu} , 
\label{eq88}
\\ 
\chi_{i0}^{\rm 1PR} (p) \simeq \chi_{0i}^{\rm 1PR} (p) \simeq & \frac{n}{m} \frac{ \omega p_{i}}{ \omega^{2} - c_{\rm T}^{2} {\bf p}^{2}}  , 
\label{eq89}
\\ 
\chi_{ij}^{\rm 1PR} (p) \simeq & \frac{nc_{\rm T}^{2}}{m} \frac{ p_{i} p_{j}}{\omega^{2} -c_{\rm T}^{2} {\bf p}^{2}} , 
\label{eq90}
\end{align} 
where $c_{\rm T}$ is the isothermal sound speed that can be found in the single-particle Green's function (\ref{eq70}). 
Here, we used a relation 
\begin{align} 
\langle - | G (p) | - \rangle  
=
\frac{4 n_{0} m c_{\rm T}^{2}}{n} \frac{1}{\omega^{2}  - c_{\rm T}^{2} {\bf p}^{2}}, 
\label{eq97}
\end{align} 
which is conveniently obtained from (\ref{eq70}).  
From Eqs. (\ref{eq70}) and (\ref{eq97}), the effect of the Nepomnyashchii--Nepomnyashchii identity and the infrared divergence of the longitudinal susceptibility $G_{\parallel} = - \langle + | G | + \rangle/4 = 1/ 4 \Sigma_{12} (0)$, which comes from the second term of (\ref{eq70}), are clearly found to be irrelevant to the density and current correlation functions.

The correlation functions can be obtained from the 1PI parts \eqref{eq84}-\eqref{eq87} and the 1PR parts \eqref{eq88}-\eqref{eq90}, given by~\cite{Gavoret:1964gv} 
\begin{align}
\chi_{00}^{} (p) \simeq & 
\frac{n}{m} \frac{{\bf p}^{2}}{ \omega^{2} - c_{\rm T}^{2} {\bf p}^{2} }, 
\label{eq79}
\\
\chi_{0i} (p) \simeq \chi_{i0} (p) \simeq & \frac{n}{m} \frac{ \omega p_{i}}{ \omega^{2} - c_{\rm T}^{2} {\bf p}^{2}} , 
\label{eq80}
\\ 
\chi_{ij} (p) \simeq & \frac{nc_{\rm T}^{2}}{m} \frac{ p_{i} p_{j}}{ \omega^{2} -c_{\rm T}^{2} {\bf p}^{2}}. 
\label{eq81}
\end{align} 
The density-density correlation functions (\ref{eq79}) satisfies the compressibility zero-frequency sum-rule, giving the form~\cite{Hohenberg:1965,griffin1993excitations} 
\begin{align}
\lim\limits_{{\bf p} \rightarrow {\bf 0}} \chi_{00} (0, {\bf p}) = - \frac{n}{mc_{\rm T}^{2}}. 
\label{CZFSR}
\end{align} 
The compressibility zero-frequency sum-rule is exhausted by the 1PI part. 
In the low-energy limit, the 1PI and 1PR parts of the density-density correlation function in (\ref{eq84}) and (\ref{eq88}) behave as 
\begin{align}
\lim\limits_{{\bf p} \rightarrow {\bf 0}} \chi_{00}^{\rm 1PI} (0, {\bf p}) = - \frac{n}{mc_{\rm T}^{2}}, 
\quad 
\lim\limits_{{\bf p} \rightarrow {\bf 0}} \chi_{00}^{\rm 1PR} (0, {\bf p}) = 0. 
\label{NNPiRnmc2}
\end{align} 
The 1PR part vanishes in the static and low-momentum limits, and does not contribute to the compressibility zero-frequency sum-rule.

The leading term of the single-particle Green's function~\eqref{eq70} and the density and current correlation functions~\eqref{eq79}-\eqref{eq81} share the pole, which provides the phonon excitations. Because of the presence of the BEC, the two-particle Green's function involves the single-particle Green's function as in the 1PR part~\eqref{eq21}. This contribution directly involves the single-particle property to the density correlation function. 
Since the self-energy in the single-particle Green's function can be related to thermodynamic quantities as discussed in this section, 
which can provides the phonon dispersion relation with the isothermal sound speed, 
the density correlation function can consistently describe the sound mode, the speed of which is defined in terms of the macroscopic compressibility. 
The paper by Huang and Klein~\cite{Huang:1964bc} also provides a useful discussion about the phonon mode in BEC. 

The single-particle excitation is also related to the superfluidity. 
An interesting relation between them owes to the Josephson sum-rule~\cite{JOSEPHSON1966608,Holzmann:2007baa}, given by 
\begin{align}
\rho_{\rm s} = - \lim\limits_{{\bf p} \to 0} \frac{m^2 n_0 }{{\bf p}^2 G_{11} (0, {\bf p})}, 
\label{eq:JosephsonSumRule}
\end{align}
where $\rho_{\rm s}$ is the superfluid mass density. By using Eq.~\eqref{eq70} as well as the relation $\Delta \Sigma (p) \gg {\bf p}^2$ in the small momentum regime, we find the relation 
\begin{align}
\rho_{\rm s} = m n, 
\label{rhos}
\end{align}
which indicates that the superfluid mass density is exactly the total mass density at $T = 0$.

The current-current response function can be decomposed into the longitudinal and transverse response functions, given by~\cite{pitaevskii2016bose,forster2018hydrodynamic,ueda2010fundamentals} 
\begin{align}
\chi_{ij} (p) = \frac{p_i p_j}{{\bf p}^2} \chi_{\rm L} (p) + \left ( \delta_{ij} - \frac{p_i p_j}{{\bf p}^2} \right ) \chi_{\rm T} (p). 
\end{align}
These longitudinal and transverse response functions are extracted from the relations~\cite{ueda2010fundamentals}  
\begin{align}
\chi_{\rm L} (p) = & \sum\limits_{i,j} \frac{p_i p_j}{{\bf p}^2} \chi_{ij} (p), 
\label{chiL}
\\
\chi_{\rm T} (p) = & \frac{1}{2} \sum\limits_{i,j} \left ( \delta_{i,j} - \frac{p_i p_j}{{\bf p}^2} \right ) \chi_{ij} (p). 
\label{chiT}
\end{align} 
The longitudinal response function satisfies the $f$-sum rule $\chi_{\rm L} ({\bf p}\to 0, 0) = - n/m$ and the transverse response function provides the normal fluid density $n_{\rm n}$, given by $\chi_{\rm T} ({\bf p} \to 0, 0) = - n_{\rm n}/m$~\cite{pitaevskii2016bose,forster2018hydrodynamic,ueda2010fundamentals,nozieres1990theory} . As a result, the superfluid mass density can be given by $\rho_{\rm s} = m^2 \lim\limits_{{\bf p} \to 0} [ \chi_{\rm T} ({\bf p}, 0) - \chi_{\rm L} ({\bf p}, 0)]$. 
Using the results \eqref{eq87} and \eqref{eq96}, we obtain 
\begin{align}
\lim\limits_{{\bf p} \to 0} 
\chi_{\rm L} ({\bf p}, 0) = & \lim\limits_{{\bf p}\to 0} 
\frac{n^2}{n_0} \frac{{\bf p}^2}{(2m)^2} \langle - | G ({\bf p}, 0 ) | - \rangle = - \frac{n}{m} , 
\\ 
\lim\limits_{{\bf p} \to 0} \chi_{\rm T} ({\bf p}, 0) = & 0, 
\end{align} 
which is consistent with the $f$-sum rule as well as with the fact that at $T=0$, the normal fluid density is absent and the superfluid mass density is equal to the total mass density as in Eq.~\eqref{rhos}.

\section{Experimental and theoretical studies of sound modes in superfluid}\label{SecV}


This section presents an overview of the experimental and theoretical studies of excitations in superfluid $^4$He and in BECs of ultracold atomic gases, which will be helpful to bridge both fields and to push further the study of the single-particle and collective excitations in BECs in ultracold atoms. 

The static structure and dynamic structure have been intensively and extensively studied on the superfluid liquid $^4$He experimentally~\cite{Woods1965,Pike1970,Woods1973,Woods1978,Talbot1988,Stirling1990,Fak1991}. 
The dynamic structure factor $S({\bf q}, \omega)$ consists of a sharp peak superimposed on a broad background in the superfluid $^{4}$He~\cite{Miller1962,Woods1965,Woods1978,Griffin1979,Griffin1980,Griffin1981,Talbot1988,Stirling1990,Glyde1992,Glyde1992PRB,Glyde1995}. 
The sharp peak in $S({\bf q},\omega)$ is interpreted as a collective density mode as well as a (single)-quasiparticle excitation arising from the 1PR part of the density response function $\chi_{00}^{\rm 1PR}$~\cite{Talbot1988,Stirling1990,Glyde1992,Glyde1992PRB}, where the density and single-particle responses have the same pole~\cite{Stirling1990,Glyde1992PRB,Diallo2014}. 
The broad component is interpreted as the multi-particle excitation originated from the 1PI part of the density response function $\chi_{00}^{\rm 1PI}$~\cite{Griffin1979,Stirling1990,Glyde1995,Griffin1979}.

The temperature dependence of $S({\bf q}, \omega)$ is quite different above and below $T_{\rm c}$~\cite{Talbot1988}, and abruptly changes at $T_{\rm c}$ ~\cite{Stirling1990}. 
As the temperature increased, the sharp peak broadens~\cite{Stirling1990,Glyde1992,Glyde1992PRB}, which is well described by quasiparticle-quasiparticle scattering~\cite{Stirling1990}, 
and it loses intensity~\cite{Woods1978,Talbot1988,Glyde1992,Stirling1990,Glyde1992PRB}, since the condensate density decreases, which includes the single-particle Green's function to the density response function. 

At low momentum regime, the superfluid has a single phonon mode~\cite{PhysRev.107.13,Stirling1990,Glyde1992}, whose peak is very sharp at low $T$, where the phase-space for the decay of a single phonon into two is limited~\cite{Talbot1988}. 
The sharp peak at the maxon momentum region is also interpreted as a contribution from a quasiparticle excitation~\cite{Glyde1992}. 
In the high momentum regime, the superfluid $^4$He does not support a collective density mode~\cite{Stirling1990}, and the density response function in this momentum regime broadens in the normal and superfluid phases~\cite{Manousakis1986,Stirling1990,Glyde1992}. 

The broad component is considered as multi-quasiparticle excitations with the high-energy tail, which originates from roton-roton, maxon-maxon, and maxon-roton contributions~\cite{Manousakis1986,Griffin1990}.
This broad continuum does not contain a collective mode in the superfluid phase~\cite{Stirling1990}, which starts from a finite positive energy~\cite{Miller1962}. 
The broad multiphonon component and high-frequency tail are largely temperature independent~\cite{Talbot1984PRB,Talbot1988,Stirling1990}. 

Above the critical temperature, the sharp peak phonon-maxon-roton excitation disappears from $S({\bf q},\omega)$~\cite{Woods1978,Talbot1988,Stirling1990,Glyde1992,Glyde1992PRB}, 
where the single-particle Green's function does not contribute to the density response function~\cite{Talbot1988}. 
In more detail, the sharp component disappears in the maxon and roton momentum regions, 
but the peak remains well defined in the low-momentum phonon region, which indicates the existence of a collective density mode~\cite{Stirling1990,Glyde1992}. 

The dynamic structure factor $S({\bf q},\omega)$ of the superfluid $^4$He has been also studied theoretically~\cite{Miller1962,Hohenberg1964,Cheung1971,Gotze1976,Hohenberg1976,Gotze1976B,Kang1978,Talbot1983,Talbot1984PRB,Payne1985,Manousakis1986,Fukushima1988,
Stirling1990,Griffin1990,Glyde1990,Zawadowski1992,Glyde1992,Glyde1992PRB}. 
In the Bogoliubov approximation, the density-fluctuation excitation spectrum is identical to that of the quasiparticles. 
However, this approximation gives the incorrect relation $\int_{0}^{\infty} S ({\bf q}, \omega) \omega d \omega = N_{0} {\bf q}^{2} /(2m)$, where the correct sum-rule is proportional to $N$ not to $N_0$~\cite{Miller1962}. 
Other approaches may be listed, such as the Hartree--Fock approximation and self-consistent Hartree approximation~\cite{Cheung1971}, the symmetric planer-spin model analysis explaining the light-scattering data~\cite{Hohenberg1976}, the formal expressions for the one- and two-quasiparticle excitation~\cite{Fukushima1988}, the two-roton bound states~\cite{Zawadowski1992}, and various sum-rules for the density and particle operators~\cite{Stringari1992}. 

In the theoretical framework, it can be clearly seen that the condensate plays an essential role in coupling the density excitation and the quasiparticle excitation~\cite{Stirling1990,Zawadowski1992,Glyde1992,Glyde1992PRB,Stringari1992,Nepomnyashchii1992}, where this hybridization disappears above the critical temperature~\cite{Zawadowski1992}. 
In the low momentum phonon regime, the single-particle Green's function and density response function share the pole~\cite{Gavoret:1964gv,Stringari1992}. 
Above the critical temperature, where the hybridization is absent, 
the maxon-roton peak vanishes in $S({\bf q},\omega)$, which suggests that the sharp maxon-roton intensity originates from the single-particle excitation and the BEC in the superfluid $^{4}$He~\cite{Glyde1990}. 

For the hybridization, the dielectric formalism~\cite{Griffin1973,Glyde1990,Glyde1992PRB,Fliesser2001} is an approach that fulfills the Ward identities related to the conservation of particle number and the breaking of the gauge symmetry, i.e., a conserving and gapless approach by using the continuity equation~\cite{Fliesser2001}. It gives the same pole in the single-particle Green's function and the density correlation function in the superfluid phase~\cite{Fliesser2001,Glyde1992PRB}, and the density fluctuation is coupled into the single-particle excitation though the condensate~\cite{Griffin1973}. 

The sound velocity~\cite{
Leggett1965,Pethick1966,Sunakawa1969,Kebukawa1970,Kebukawa1973,Goble1974,Prakash1977,Singh1978,Ferrell1980,Ferrell1982,Um1984} as well as the sound attenuation coefficient~\cite{Andreev1963,Khalatnikov1966,Pethick1966,Ferrell1968,Andreev1970,Kebukawa1974,Ferrell1980,Ferrell1981,Ferrell1987} are theoretically investigated, where theoretical approaches include the single-particle Green's function approach~\cite{Leggett1965,Pethick1966,Prakash1977,Singh1978,Payne1985}, the collective description theory~\cite{Sunakawa1962I,Sunakawa1962II,Sunakawa1962III,Sunakawa1969,Kebukawa1970,Nishiyama1971,Kebukawa1973,Kebukawa1974,Yamasaki1975A}, and the kinetic equation approach~\cite{Andreev1963,Andreev1970,Maris1973,Um1984}. 
Since the single-particle Green's function and density response function share the pole~\cite{Stringari1992}, the sound speed and damping are calculated from the pole of the single-particle Green's function~\cite{Leggett1965,Pethick1966,Prakash1977,Singh1978,Payne1985}. 
The finite energy spread of phonon excitations are studied by using the thermodynamic perturbation theory assuming the possibility of the three-phonon interaction~\cite{Leggett1965}. 
Using the Green's function approach, the sound speed shows the temperature dependence given in the increase as $T^{4}\ln T$~\cite{Pethick1966,Prakash1977,Singh1978} and the decrease as $T^4$~\cite{Prakash1977,Singh1978}; the damping rate shows the $T^4$-law~\cite{Pethick1966,Prakash1977,Singh1978}, which comes from the three-phonon processes~\cite{Pethick1966}.

The collective description is a theory described by the canonical collective variables, i.e., the density fluctuation and velocity operators~\cite{Sunakawa1962I,Sunakawa1962II,Sunakawa1962III,Sunakawa1969,Kebukawa1970,Nishiyama1971,Kebukawa1973,Kebukawa1974,Yamasaki1975A}, which is a divergent free approach~\cite{Sunakawa1969,Kebukawa1970,Kebukawa1973}. 
The collective description is employed to study energy spectrum~\cite{Sunakawa1969,Kebukawa1970,Nishiyama1971,Kebukawa1973}, focusing on effects of the phonon-phonon interaction~\cite{Sunakawa1969}, and phonon-roton interaction~\cite{Kebukawa1973}, which play an important role in the phonon velocity and roton minimum, and is employed to study the temperature dependence of the sound velocity and the absorption coefficient including the thermal roton effect~\cite{Kebukawa1974}.

The kinetic equation is also applied to study the sound velocity and absorption~\cite{Andreev1963,Andreev1970,Maris1973,Um1984}. 
In this approach, collisions between excitations are assumed to be not frequent in the superfluid helium at low temperatures, and thus the kinetic equation in the collisionless regime is employed. 
The sound velocity in the sufficiently low temperatures increases as $T^{4} \ln ({\rm const.}/T)$~\cite{Andreev1963}, where the constant is very small~\cite{Andreev1970}, and the absorption is reported to follow the $T^{6}$-law~\cite{Andreev1963}.

Since successful creation of the BEC in alkali atom gases~\cite{Davis1995,Anderson:1995vb}, 
the condensate excitation in ultracold gases has been intensively and extensively studied~\cite{Dalfovo1999,Leggett2001,RevModPhys.77.187}. 
Sudden modification of the trapping potential can create the local density fluctuation, and the dynamical propagation of the density fluctuation has been measured by using the phase-contrast images, where the propagation speed is consistent with the Bogoliubov theory~\cite{Andrews1997}. 
Two-photon Bragg scattering is a useful tool to study the excitation in the BEC of ultracold gases~\cite{Stenger1999}. 
The Bragg spectroscopy has been applied to measure the structure factor of the BEC in the phonon regime, the line shift and line strength of which are consistent with the results of the local density approximation~\cite{Stamper-Kurn1999}. 
The Bragg pulses have also been applied to observe the Bogoliubov transformation for a BEC~\cite{Brunello2000,Vogels:2002be}, and to reveal the wide range of the excitation spectrum from the phonon regime to the single-particle regime, which is also consistent with the Bogoliubov theory with the local density approximation~\cite{Steinhauer:2002hgb}. 
By using the Bragg spectroscopy, experiments have probed the excitation in a strongly interacting BEC~\cite{Papp:2008iu}, as well as the roton-type excitation in BECs with cavity-mediated long-range interactions~\cite{Mottl1570}, with spin-orbit couplings~\cite{Ji:2015ih}, in shaken optical lattices~\cite{Ha:2015ja}, and with dipole interactions~\cite{Petter:2019fx}. 
Recently, the sound propagation of the BEC trapped in a box trap has been intensively and extensively studied, including a uniform two-dimensional Bose gas~\cite{Ville:2018gs}, and a cylindrical box trap with tuning the atomic density~\cite{Garratt:2019ir}, which are free from the conventional restriction of the harmonic trap potential.

Through the development of the study on BECs in ultracold atoms, 
theories have been proposed~\cite{Kita:2010fv,Kita:2011gqb,Kita:2014bc,Kita:2019if} that cast doubt upon the conventional wisdom about the BEC, where those recent theories claim that the dispersion relation of the single-particle excitation is not phonon and not equal to that of the collective excitation in the low-energy and low-momentum regime, which contradicts the earlier result given by Gavoret and Nozi{\`e}res~\cite{Gavoret:1964gv}. 
It is concluded from two different approaches: the Luttinger-Ward thermodynamic functional approach ($\Phi$-derivable approximation)~\cite{Kita:2009jm,Kita:2010fv,Kita:2011gqb,Kita:2014bc} and a functional renormalization group approach~\cite{Kita:2019if,Kita:2019}. 

The Luttinger-Ward thermodynamic functional approach~\cite{Kita:2009jm,Kita:2010fv,Kita:2011gqb,Kita:2014bc} is useful for considering the theory satisfying the Noether's theorem and the Goldstone's theorem, which may cure the so-called conserving-gapless dilemma~\cite{Hohenberg:1965,Griffin:1996tj,Yukalov:2008exb,Yukalov:2011ef}. The papers~\cite{Kita:2011gqb,Kita:2014bc} are concluded that the self-energy contribution should be one-particle reducible (1PR), because the 1PR contribution cures the conserving-gapless dilemma. 
As a result, the two-particles Green's function has the pole showing the collective sound mode; on the other hand, the single-particle Green's function provides a bubbling mode with a considerable decay rate rather than the sound mode, which results in no well-defined quasiparticle in BECs~\cite{Kita:2011gqb}. However, in general, in the case where the self-energy contribution is included to the Green's function through the Dyson-Beliaev equation, the one-particle irreducible part should be employed. Otherwise, multi-counting of diagrammatic contribution emerges in the full Green's function. In this regard, even if the 1PR approximation may avoid the conserving-gapless dilemma, it provides a problem, namely, the trilemma among conserving, gapless and 1PR approximation in the BEC theory.

By using the exact renormalization-group technique~\cite{kopietz2010introduction,Sinner:2010jy}, 
the study~\cite{Kita:2019if} concluded that the one-particle density matrix approaches asymptotically the condensate density as $1/r^{d-2+\eta}$ with an anomalous dimension $\eta > 0$, which gives the single-particle Green's function $G_{11} \propto 1/{\bf p}^{2-\eta}$ in the low-momentum regime. As a result, the paper~\cite{Kita:2019if} claimed that a three-dimensional BEC at $T=0$ does not have the Bogoliubov phonon mode. The behavior $G_{11} \propto 1/{\bf p}^{2-\eta}$, however, provides an unphysical situation, which gives the superfluid density $\rho_{\rm s}$ being infinity according to the Josephson sum-rule~\eqref{eq:JosephsonSumRule}~\cite{JOSEPHSON1966608,Holzmann:2007baa,ueda2010fundamentals}. The anomalous dimension $\eta > 0$ also violates the Bogoliubov operator inequality $-G_{11} (0,{\bf p}) \geq m n_0 /(n{\bf p}^2)$~\cite{Baym1968,Holzmann:2003eu,Yukalov:2008exb,Yukalov:2016ft,forster2018hydrodynamic,pitaevskii2016bose,ueda2010fundamentals}. 
According to this Bogoliubov operator inequality, the relation $\eta = 0$ should hold for $T < T_{\rm c}$~\cite{Holzmann:2003eu}. 
The possibility of the anomalous dimension $\eta >0$ emerges only in the case at precisely $T_{\rm c}$, where the correlation length diverges~\cite{Holzmann:2003eu}. 

Other approach, the extension of Bijl--Feynman formula, also provides the result that the energy spectrum of the single-particle excitation is distinct from that of the collective excitation, and the lifetime of the quasiparticles remains finite even in the long-wavelength limit~\cite{Tsutsumi2016}. In this respect, the Josephson sum-rule~\eqref{eq:JosephsonSumRule} and the Bogoliubov operator inequality~\cite{Baym1968,Holzmann:2003eu,Yukalov:2008exb,Yukalov:2016ft,forster2018hydrodynamic} could be useful criteria for the result contradictory to the conventional wisdom about the single-particle excitation and the collective excitation in BECs.

\section{Density response function in random phase approximation}\label{SecVI}

The matrix formalism is a useful tool to develop many-body theories, such as the random phase approximation (RPA), for studying many-body effects as well as the density-density correlation function in BECs. 
The same idea of the matrix formalism for the BEC may be found in the study of an effective roton-maxon interaction in liquid He II~\cite{Fukushima1988}. 
In the BEC phase, the density-density correlation function is constructed from the sum of the 1PI and 1PR parts as in (\ref{eq36}), i.e., 
$\chi_{00} = \chi_{00}^{\rm 1PI} + \chi_{00}^{\rm 1PR} $.  
In the following, we omit the subscript describing the density vertex $\mu = \nu = 0$ in the polarization function $\chi$ as well as the density vertices $\Upsilon$ and $\Upsilon^\dag$, for simplicity. 

We consider the 2PI parts $I(p,p';q)$, $J(p;q)$ and $J^\dag (p;q)$ introduced in Sec.~\ref{SecII} as the simplest contributions, given by 
\begin{align}
I (p, p'; q) = &U + \frac{1}{2} | f_0 \rangle U \langle f_0 | \equiv \hat U , 
\\
J (p,q) = & - \sqrt{-1} \hat U ( \mathcal{G}_{1/2} + \hat X \mathcal{G}_{1/2} ) ,
\\
J^\dag (p,q) = & - \sqrt{-1} ( \mathcal{G}_{1/2}^\dag + \mathcal{G}_{1/2}^\dag \hat X)  \hat U,
\end{align}
where the first and second terms in $\hat U$ provide the Hatree and Fock contributions in the present matrix formalism, respectively. 
By assuming the momentum and frequency-dependence of the four and three point vertices as 
$\Gamma(p, p'; q) = \Gamma (q)$, $P(p;q) = P(q)$, and $P^\dag(p;q) = P^\dag(q)$, we construct the random phase approximation by using Eqs. \eqref{eq23}, \eqref{eq27} and \eqref{eq28}, giving the forms 
\begin{align}
\Gamma(q) = & \frac{1}{1 - \hat U \Pi (q)} \hat U , 
\\ 
P (q) = &  - \sqrt{-1} \Gamma (q) (1 + \hat X ) {\mathcal G}_{1/2},
\\  
P^\dag (q) = &- \sqrt{-1} {\mathcal G}_{1/2}^\dag ( 1 + \hat X) \Gamma (q) , 
\end{align}  
where $\Pi (q) = - T \sum_{p} K_0 (p;q)$. 
The 1PI part of the density correlation functions and the density vertices are also given by 
\begin{align}
\chi^{\rm 1PI} (q) = & \frac{1}{2} \langle f_0 | [ { \Pi} (q) + { \Pi} (q) { \Gamma} (q) { \Pi} (q) ] |f_0 \rangle, 
\label{eqnew59}
\\ 
\Upsilon (q) = & - \sqrt{-1} \left [ G_{1/2}
+ \frac{1}{2} {\mathcal G}_{1/2}^\dag (1  + \hat X )  \Gamma (q) \Pi (q) | f_0 \rangle 
\right ] , 
\label{eqnew60}
\\ 
\Upsilon^\dag (q) = & - \sqrt{-1} \left [ G_{1/2}^\dag
+ \frac{1}{2} \langle f_0 | \Pi (q) \Gamma (q) (1+ \hat X ) {\mathcal G}_{1/2}  \right ]. 
\label{eqnew61}
\end{align} 
By using the relations such as $G_{22}(p) = G_{11} (-p)$ as well as $G_{12}(p) = G_{12}(-p)$, the polarization function can be reduced into 
\begin{align}
\Pi(q) = & 
\begin{pmatrix}
\Pi_{11} (q) & \Pi_{12} (q) & \Pi_{12} (q) & \Pi_{14} (q) \\
\Pi_{12} (q) & \Pi_{22} (q) & \Pi_{14} (q) & \Pi_{12}^{*} (q) \\
\Pi_{12} (q) & \Pi_{14} (q) & \Pi_{22} (q) & \Pi_{12}^{*} (q) \\
\Pi_{14} (q) & \Pi_{12}^{*} (q) & \Pi_{12}^{*} (q)& \Pi_{11}^{*}  (q)
\end{pmatrix}. 
\label{RPAPiMatrix}
\end{align}

The four point vertex in this approximation can be conveniently decomposed into the $T$-matrix ${\mathcal T}(q)$ given by the ladder type diagrams and the effective interaction $U_{\rm eff}(q)$ including the density fluctuation, given by 
\begin{align}
\Gamma (q) = & 
{\mathcal T} (q) 
+ \frac{1}{2} | f_0 \rangle U_{\rm eff} (q) \langle f_0 | 
+ \frac{1}{2} \boldsymbol{\mathit \gamma}^{} (q) U_{\rm eff} (q) \boldsymbol{\mathit \gamma}^{\dag}  (q) 
\nonumber 
\\ 
& 
+ \frac{1}{2}  \boldsymbol{\mathit \gamma}^{} (q) U_{\rm eff} (q) \langle f_0 | 
+ \frac{1}{2} | f_0 \rangle U_{\rm eff} (q) \boldsymbol{\mathit \gamma}^{\dag} (q) , 
\end{align} 
where $\boldsymbol{\mathit \gamma} (q)  = {\mathcal T} (q) { \Pi} (q) |f_0 \rangle$, $\boldsymbol{\mathit \gamma}^{\dag} (q) = \langle f_0 | { \Pi} (q) {\mathcal T} (q)$ and 
\begin{align}
{\mathcal T} (q) = & \frac{U}{1 - U { \Pi}(q)},
\label{RPABetheSalpeter0}
\\ 
U_{\rm eff} (q) = & \frac{U}{ 1 - U \chi_{\rm R} (q) }. 
\label{RPABetheSalpeter}
\end{align}  
Here, $\chi_{\rm R} (q)$ is the regular part of the density-density correlation function including the vertex correction, giving the form 
\begin{align}
\chi_{\rm R} (q) = & \frac{1}{2} \langle f_0 | [ { \Pi} (q) + { \Pi} (q) {\mathcal T} (q) { \Pi} (q) ] |f_0 \rangle. 
\end{align}
The 1PI part of the density correlation function (\ref{eqnew59}), and the density vertices (\ref{eqnew60}) and (\ref{eqnew61}) are also reduced into 
\begin{align}
\chi^{\rm 1PI}(q) = & \frac{ \chi_{\rm R}(q)  }{ 1 - U \chi_{\rm R}(q)}. 
\label{Eq259}
\\ 
\Upsilon (q) = & - \sqrt{-1} \left [ G_{1/2} + \frac{1}{2} {\mathcal G}_{1/2}^{\dag} (1 +  \hat X ) \boldsymbol{\mathit \gamma}^{} (q) \right ] 
A (q) , 
\label{eq76g}
\\
\Upsilon^{\dag} (q) = & - \sqrt{-1} A(q) \left [ G_{1/2}^{\dag} 
+ \frac{1}{2} \boldsymbol{\mathit \gamma}^{\dag} (q) ( 1 + \hat X) {\mathcal G}_{1/2} \right ], 
\label{eq77g}
\end{align}
where $A (q) = [1 + U_{\rm eff} (q)  \chi_{\rm R} (q)  ]$. 
It can be clearly seen from the present formalism that the condensate plays an essential role in coupling the density excitation and the quasiparticle excitation as in Refs.~\cite{Stirling1990,Zawadowski1992,Glyde1992,Glyde1992PRB,Stringari1992,Nepomnyashchii1992}, where this hybridization disappears above the critical temperature~\cite{Zawadowski1992}. 
Note that because of the relation (\ref{RPAPiMatrix}), 
five elements ${\mathcal T}_{11, 12, 14, 22, 23}$ are needed to construct the $T$-matrix ${\mathcal T}$, which is given by 
\begin{align}
{\mathcal T}(q) = & 
\begin{pmatrix}
{\mathcal T}_{11} (q) & {\mathcal T}_{12} (q) & {\mathcal T}_{12} (q) & {\mathcal T}_{14} (q) \\ 
{\mathcal T}_{12} (q) & {\mathcal T}_{22} (q) & {\mathcal T}_{23} (q) & {\mathcal T}_{12}^* (q) \\ 
{\mathcal T}_{12} (q) & {\mathcal T}_{23} (q) & {\mathcal T}_{22} (q) & {\mathcal T}_{12}^* (q) \\ 
{\mathcal T}_{14} (q) & {\mathcal T}_{12}^{*} (q) & {\mathcal T}_{12}^* (q) & {\mathcal T}_{11}^{*} (q) 
\end{pmatrix}. 
\label{GammaMatrix}
\end{align} 

We take the following bare part of the two-particle Green's function 
\begin{align}
K_0 (p; q) = g (p+q) \otimes g(-p), 
\end{align} 
where $g_{} (p)$ is the single-particle Green's function, given by 
\begin{align}
g(p) = \left \{ 
\begin{array}{ll}
\displaystyle{ \frac{1}{ i \omega_{n} \sigma_{3} - \xi_{\bf p} - U n_{0} \sigma_{1} } } \quad (T \leq T_{\rm c})
\\[10pt]
\displaystyle{ \frac{1}{ i \omega_{n} \sigma_{3} - \varepsilon_{{\bf p}} + \mu - \Sigma_{11} (0) } } \quad (T \geq T_{\rm c}) , 
\end{array}
\right . 
\end{align} 
where $\xi_{\rm p} = \varepsilon_{\bf p} + U n_{0}$. 
At $T \leq T_{\rm c}$, we employed the Hartree--Fock--Bogoliubov--Popov (Shohno) approximation~\cite{popov2001functional,Griffin:1996tj,Popov:1964wj,Popov:1965wt,SHI19981,Shohno:1964}. 
At $T \geq T_{\rm c}$, the effective chemical potential is taken to be $\mu - \Sigma_{11} (0)$, since the Green's function $g$ has a pole of a gapless dispersion law at the critical temperature $T_{\rm c}$, with satisfying the Hugenholtz-Pines relation $\mu = \Sigma_{11} (0)$. 
The detailed expressions of the polarization function $\Pi_{11,12,14,22}$ in this approximation are summarized in Appendix~\ref{AppendixD}.

Above the critical temperature, the 1PI part of the density-density correlation function is given by 
\begin{align}  
\chi^{\rm 1PI} (q) = \frac{\Pi_{22} (q) }{1-2 U\Pi_{22} (q) }, 
\label{eq98}
\end{align} 
because $g_{12} (p) = \Pi_{12,14} (p) = 0$ at $T \geq T_{\rm c}$. 
At the same temperature regime, the regular part is given by 
\begin{align} 
\chi_{\rm R} (q) = & \frac{\Pi_{22} (q) }{1-U\Pi_{22} (q) }. 
\end{align} 
The $T$-matrix ${\mathcal T}$ at $T \geq T_{\rm c}$ has a diagonal form, whose matrix elements are given by 
\begin{align}
{\mathcal T}_{11(22)} (q) = \frac{U}{1-U \Pi_{11(22)} (q) }. 
\end{align}

For the single-particle Green's function $G(p)$, 
we include many-body effects to the self-energy by using the RPA for focusing on density fluctuations, which is given by 
\begin{align}
\Sigma_{11} (p) = & (n_0+ \tilde n)U_{\rm eff} (0) 
\nonumber \\ & 
+ n_{0} U_{\rm eff} (p) 
- T \sum\limits_{q} U_{\rm eff} (q) g_{11} (p-q), 
\label{eq27s} 
\\
\Sigma_{12} (p) = & n_{0} U_{\rm eff} (p) , 
\label{eq28s} 
\end{align} 
where $\tilde n= - T \sum_{p} g_{11} (p)$.

The density vertex $\Upsilon$ in the RPA given in \eqref{eq76g} does not satisfy the zero-frequency density vertex identity $\Upsilon (0) = 0$. 
In the static and low-momentum limit, the density vertices given in (\ref{eq76g}) and (\ref{eq77g}) are reduced to 
\begin{align} 
\begin{pmatrix} \Upsilon (0)  \\ \Upsilon^\dag (0) \end{pmatrix}
= & \frac{\sqrt {n_{0}}}{1 - U \chi_{\rm R}(0)} \frac{2 \Gamma' (0)}{U}  
\begin{pmatrix} |+ \rangle  \\ \langle + |  \end{pmatrix}
\label{eq155} 
\end{align}
where 
\begin{align}
\chi_{\rm R} (0) 
= - \frac{1}{U} \frac{1 - U \Pi' (0) }
{ 2 - U \Pi' (0) }, 
\quad 
\Gamma_{}' (0) =   \frac{U}{2 - U \Pi' (0)}, 
\end{align} 
with 
\begin{align} 
\Pi' (q) = & \Pi_{11} (q) + \Pi_{22} (q) + 2 \Pi_{14} (q) + 4 \Pi_{12} (q) . 
\label{eq119}
\end{align} 
Each polarization function $\Pi_{ij}$ exhibits an infrared divergence. 
For example, in the three dimensional system at $T\neq 0$, the polarization functions exhibit the infrared divergence, giving the form $\Pi_{11,12,22,14} ({\bf p}, 0) \propto 1/|{\bf p}|$ at small ${\bf p}$~\cite{Watabe:2014gv}.
Because of a relation $g_{11} (p)=-g_{12}(p)$ in the low-energy limit, 
the following exact relation holds:  
\begin{align}
\lim_{{\bf p} \to0} \Pi_{11,22,14} (i \omega_{n} = 0, {\bf p}) = - \lim_{p\to0} \Pi_{12} (i \omega_{n} = 0, {\bf p}). 
\label{120}
\end{align} 
All the infrared divergences are thus canceled out each other in $\Pi'$, and then the function $\Pi' (0)$ converges at $T < T_{\rm c}$. 
By using (\ref{gapless Hartree-Fock-BogoliubovPi11fp}), (\ref{gapless Hartree-Fock-BogoliubovPi12fp}), (\ref{gapless Hartree-Fock-BogoliubovPi14fp}) as well as (\ref{gapless Hartree-Fock-BogoliubovPi22fp}), we have its explicit form given by 
\begin{align}
 \lim\limits_{{\bf q} \rightarrow {\bf 0}}\Pi' (0,{\bf q}) 
= & 
\sum\limits_{\bf p} 
\frac{ \varepsilon_{\bf p}^{2} }{ E_{\bf p}^{2} } 
\left ( 
\frac{\partial n_{\bf p}}{ \partial E_{\bf p}} 
- 
\frac{1 +2  n_{\bf p}}{2 E_{\bf p}}  
\right ) , 
\label{eq88Pi}
\end{align} 
where $n_{\bf p}$ is the Bose-distribution function $n_{\bf p} = 1/  [ \exp{(E_{\bf p} / T)}- 1 ]$ with $ E_{\bf p} = \sqrt{\varepsilon_{\bf p} (\varepsilon_{\bf p} + 2 U_{} n_{0} )}$. 
At $T < T_{\rm c}$, therefore, the density vertex parts in (\ref{eq155}) provide $\Upsilon (0) \neq 0$. 

This problem can be avoided by adopting the simplified regular part of the density-density correlation function that does not include the vertex correction, giving the form 
\begin{align}
\chi_{\rm R}^{\rm s} (q) = \frac{1}{2} \langle f_{0} | \Pi (q) | f_{0} \rangle. 
\label{Eq259simpler}
\end{align} 
Using this simplified version, 
we may take a variant of the density vertices $\Upsilon^{\rm s} (q)$ and $\Upsilon^{{\rm s} \dag} (q)$, which are given by replacing $A(q)$ in Eqs. \eqref{eq76g} and \eqref{eq77g} with $A^{\rm s} (q) = 1 + U_{\rm eff}^{\rm s} (q)  \chi_{\rm R}^{\rm s} (q)$, 
where 
\begin{align}
U_{\rm eff}^{\rm s} (q) =  \frac{U}{ 1 - U \chi_{\rm R}^{\rm s} (q) }. 
\label{RPABetheSalpeters}
\end{align}  
In the low-energy limit, the simplified density vertex $\Upsilon^{\rm s}$ is reduced to 
\begin{align} 
\begin{pmatrix} \Upsilon^{\rm s} (0)  \\ \Upsilon^{{\rm s} \dag} (0) \end{pmatrix}
= & \frac{\sqrt {n_{0}}}{1 - U \chi_{\rm R}^{\rm s}(0)} \frac{2 \Gamma' (0)}{U}  
\begin{pmatrix} |+ \rangle  \\ \langle + |  \end{pmatrix}. 
\label{eq123Us}
\end{align}
Since the simplified regular part is given by $\chi_{\rm R}^{\rm s} (0) = \chi_{22} (0) + \chi_{14} (0)$, which shows the infrared divergence, 
the simplified density vertex $\Upsilon^{\rm s}$ satisfies the identity $\Upsilon^{\rm s} (0) = 0$.

According to the same reason, the off-diagonal self-energy (\ref{eq28s}) does not satisfy the Nepomnyashchii--Nepomnyashchii identity $\Sigma_{12} (0) = 0$. This problem is also avoided by replacing the effective interaction $U_{\rm eff} $ with $U_{\rm eff}^{\rm s}$ in the off-diagonal self-energy (\ref{eq28s}), because the infrared divergence of $\chi_{\rm R}^{\rm s}$ provides $U_{\rm eff}^{\rm s} (0) = 0$. 
As a result, the off-diagonal self-energy $\Sigma_{12} = n_{0} U_{\rm eff}^{\rm s} (p)$ is one of the candidates to satisfy the Nepomnyashchii--Nepomnayshchii identity~\cite{Nepomnyashchii:1978wb,Watabe:2013hw}. 
Other approaches that satisfies the Nepomnyashchii--Nepomnayshchii identity have been also discussed, including the description in terms of the hydrodynamic variables~\cite{Popov1972,Popov:1979vk,popov2001functional,Dupuis:2011gf,Dupuis2011,Stoof:2013dva,Watabe:2014gv}, the renormalization group approach~\cite{Bijlsma1996,Sinner:2009bm,Sinner:2010jy,Dupuis:2011gf,Stoof:2013dva}, the large-$N$ expansion~\cite{Hryhorchak:2018fe,Dupuis:2011gf} 
and the division approach into singular and nonsingular self-energies~\cite{Watabe:2014gv}.

\section{Density and single-particle spectral function}\label{SecVII}

\begin{figure*}[tb]
\begin{center}
\includegraphics[width=\textwidth]{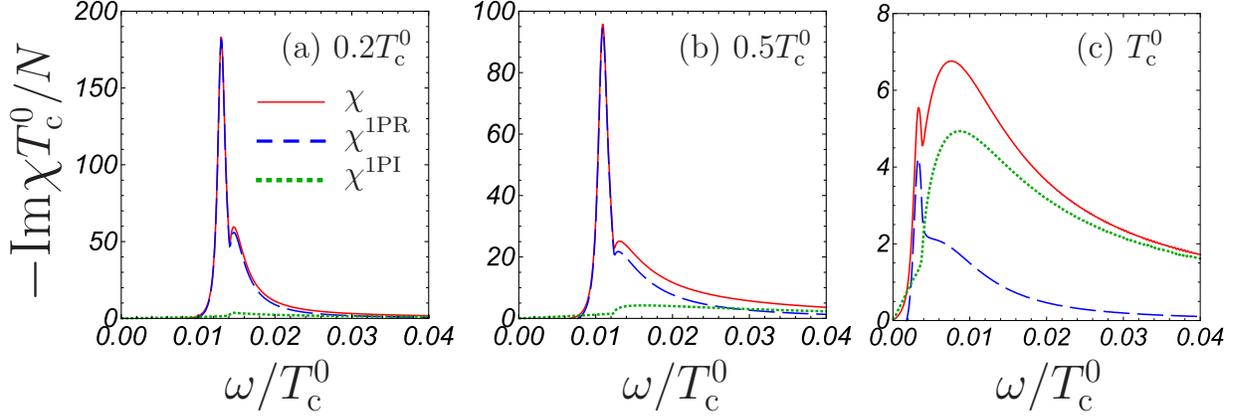}
\end{center}
\caption{Density response function at temperature (a) $0.2T_{\rm c}^{0}$, (b) $0.5T_{\rm c}^{0}$, and (c) $T_{\rm c}^{0}$, where $T_{\rm c}^{0}$ is the critical temperature of an ideal Bose gas. 
We used the self-energies in (\ref{eq27s}) and (\ref{eq28s}), and the density vertex in (\ref{eq76g}) and (\ref{eq77g}), with the effective interaction (\ref{RPABetheSalpeter}) including the vertex correction. 
We also used the 1PI part in (\ref{Eq259}) also including the vertex correction. 
The critical temperature is given by $T_{\rm c} /T_{\rm c}^{0} \simeq 1 + 1.9 an^{1/3}$ at the gas parameter $an^{1/3} = 10^{-2}$. 
We take the momentum $q = 0.05q_{0}$, where $T_{\rm c}^{0} \equiv q_{0}^{2} / (2m)$. 
}
\label{fig_graph1.fig}
\end{figure*} 

This section serves as the study of the density response function and the single-particle spectral function in the BEC by using the formalism developed in the previous section. 
The condensate density is calculated as a function of temperature, 
by solving the particle number equation with the non-condensate density \eqref{eq17}, where below $T_{\rm c}$, the chemical potential satisfies the Hugenholtz-Pines relation, and the self-energies are given in \eqref{eq27s} and \eqref{eq28s}. 
We performed the analytic continuation based on Refs.~\cite{PhysRevB.37.4965,PhysRevB.52.12720}. 

At the low temperature regime ($0.2 T_{\rm c}^{0}$), where $T_{\rm c}^{0}$ is the critical temperature of an ideal Bose gas, the sharp peak emerges with the satellite structure in the density response function $\chi$ (Fig.~\ref{fig_graph1.fig}(a)). 
Since the 1PI part $\chi^{\rm 1PI}$ is found to be negligibly small compared with the 1PR part $\chi^{\rm 1PR}$, the satellite peak is mainly originated from $\chi^{\rm 1PR}$ part at the low temperature. 
This is stark contrast to the case of the multi-particle excitation in the superfluid $^4$He, which provides the significant broad peak. 
The multi-particle excitation in the superfluid $^4$He is originated from the roton-roton, maxon-maxon, and roton-maxon scattering and their bound states. Since the dispersion relation of the quasiparticle has extremum at the roton and maxon region, those provides the very large density of states owing to the van Hove singularity. 
This effect leads the pronounced contribution of the 1PI part to the density response function. 
In the present case without roton and maxon excitations, however, the satellite peak is originated from the 1PR part. 
For increasing temperature, the contribution from the 1PI part is enhanced (Fig.~\ref{fig_graph1.fig}(b)), and the density response function is mainly organized by the 1PI part close to $T_{\rm c}$ (Fig.~\ref{fig_graph1.fig}(c)). 
Although the main structure of $\chi$ in Fig.~\ref{fig_graph1.fig}(c) is the broad peak with a tail in the high frequency side, one can see the small sharp peak structure at the low frequency side, which is originated from the 1PR part. 
Very close to the critical temperature, the 1PR part does not show the main contribution to the density response function, because the density vertex $\Upsilon$ proportional to $\sqrt{n_{0}}$ is small.

\begin{figure}
\begin{center}
\includegraphics[width=8cm]{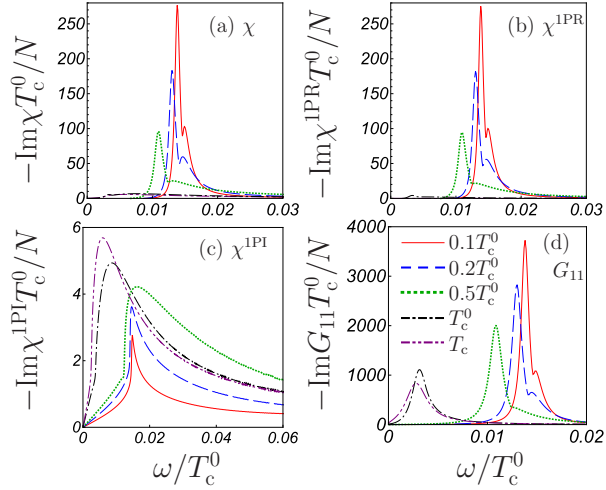} 
\end{center}
\caption{
Structure of the response functions at  $an^{1/3} = 10^{-2}$ and $q = 0.05q_{0}$. 
(a) The total density response function $\chi$, (b) the 1PR part $\chi^{\rm 1PR}$, (c) the 1PI part $\chi^{\rm 1PI}$, and (d) the single-particle Green's function $G_{11}$. 
We used the same Feynman diagrams as used in Fig.~\ref{fig_graph1.fig}. 
} 
\label{fig_graph2.fig}
\end{figure}

The temperature dependences of each contribution to $\chi$ are summarized in Fig.~\ref{fig_graph2.fig}. 
The density response function gives the striking sharp peak with the satellite structure, but at the intermediate temperature, the peak strength becomes weak and the satellite peak structure changes into the tail structure (Fig.~\ref{fig_graph2.fig}(a)). 
The intensity of the density response function at the critical temperature is quite small compared with the case at the low temperature. 
The 1PR part shows the similar behavior to the total density response function $\chi$; however, the 1PR part vanishes at the critical temperature (Fig.~\ref{fig_graph2.fig}(b)). 
The 1PI part exhibits the striking sharp structure with a broad tail at $0.1T_{\rm c}^{0}$; on the other hand, as the temperature increase, this sharpness vanishes with the growth of the intensity (Fig.~\ref{fig_graph2.fig}(c)). 
The spectral function of the single-particle excitation is also shown in Fig.~\ref{fig_graph2.fig}(d). 
The structure of $G_{11}$ in the low temperature regime provides the sharp peak with a small satellite peak, which is the same behavior as the density response functions $\chi$ and $\chi^{\rm 1PR}$. 
However, at high temperature such as $T_{\rm c}^{0}$ and $T_{\rm c}^{}$, 
the satellite peak disappears, where the intensity of $-{\rm Im}G_{11}$ remains the same order as those in the low-temperature case, which is in contrast to the case for the density response function. 
In the density response function $\chi$, the peak of the 1PR part emergent from the single-particle excitation is suppressed by the density vertex $\Upsilon$ proportional to $\sqrt{n_0}$.

\begin{figure}[tb]
\begin{center}
\includegraphics[width=8cm]{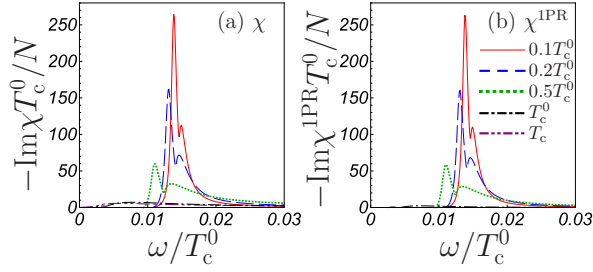} 
\end{center}
\caption{
Structure of the response functions at $an^{1/3} = 10^{-2}$ and $q = 0.05q_{0}$. 
(a) The total density response function $\chi$, (b) the 1PR part $\chi^{\rm 1PR}$. 
We used the same structure of the self-energies and the 1PI part as in Fig.~\ref{fig_graph1.fig}.  
For the density vertex, we employed $\Upsilon^{\rm s}$ and $\Upsilon^{{\rm s} \dag}$, which satisfies the zero-frequency density vertex identity $\Upsilon^{\rm s} (0) = 0$. 
} 
\label{fig_graph3.fig}
\end{figure}

\begin{figure*}[tb]
\begin{center}
\includegraphics[width=\textwidth]{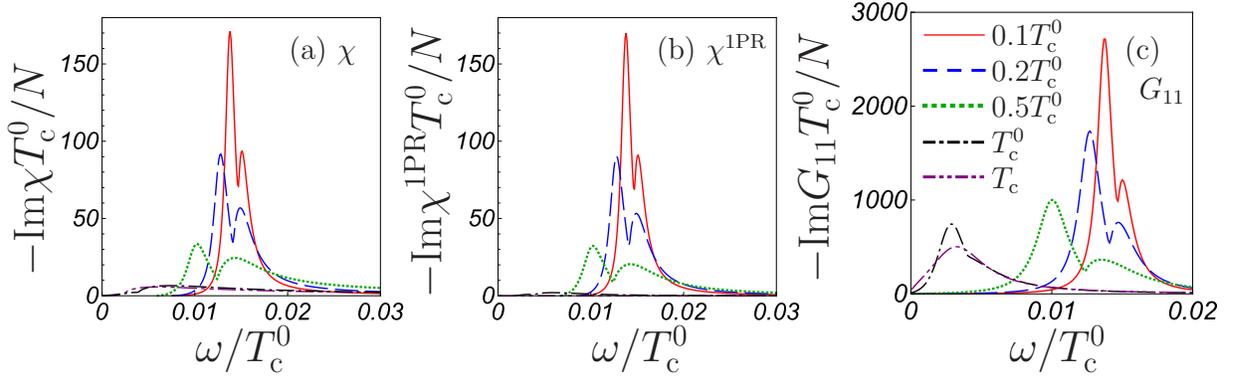}
\end{center}
\caption{
Structure of the response functions at $an^{1/3} = 10^{-2}$ and $q = 0.05q_{0}$. 
(a) The total density response function $\chi$, (b) the 1PR part $\chi^{\rm 1PR}$, and (c) the single-particle Green's function $G_{11}$.
We used the self-energy satisfying the Nepomnyashchii--Nepomnyashchii identity $\Sigma_{12}(0) = 0$, where $U_{\rm eff}$ in (\ref{eq27s}) and (\ref{eq28s}) is replaced with $U_{\rm eff}^{\rm s}$ in (\ref{RPABetheSalpeters}). 
For the density vertex, we employed $\Upsilon^{\rm s}$ and $\Upsilon^{{\rm s} \dag}$. 
We used the 1PI part in (\ref{Eq259}) including the vertex correction. 
}
\label{fig_graph4.fig}
\end{figure*}

In Figs.~\ref{fig_graph1.fig} and \ref{fig_graph2.fig}, we have discussed the structure of the density response function and the single-particle spectral function by using the self-energies (\ref{eq27s}) and (\ref{eq28s}) and the density vertex (\ref{eq76g}) and (\ref{eq77g}) both including the vertex correction. 
These qualitative features do not change in the case where the vertex correction is eliminated. Figure~\ref{fig_graph3.fig} shows the results with the density vertices $\Upsilon^{\rm s}$ and $\Upsilon^{{\rm s} \dag}$ satisfying the identity $\Upsilon^{\rm s} (0) = 0$, where the self-energy contribution is still given by  (\ref{eq27s}) and (\ref{eq28s}). Figure~\ref{fig_graph4.fig} shows the results with the density vertex $\Upsilon^{\rm s}$ and $\Upsilon^{{\rm s} \dag}$ as well as the self-energy contribution with the use of the effective interaction (\ref{RPABetheSalpeters}), which satisfies the Nepomnyashchii--Nepomnyashchii identity $\Sigma_{12} (0) = 0$~\cite{Watabe:2018id}. 
Although the satellite peaks without the vertex correction in Figs.~\ref{fig_graph3.fig} and~\ref{fig_graph4.fig} are very slightly enhanced compared with the result in Fig.~\ref{fig_graph1.fig}, the qualitative features remain the same.

Above the critical temperature, the density response function is exhausted by the 1PI part, where the 1PR part is absent since $\Upsilon = 0$. 
By using the RPA, we found that the density response function at $T > T_{\rm c}$ has qualitatively the same structure at $T = T_{\rm c}$, 
where a broad structure emerges and very long-lived collective excitations are absent. 
This is because the random phase approximation describes collisionless modes, and does not describe the hydrodynamic mode. 
In this sense, this result indicates that there is no long-lived collisionless sound modes in a normal Bose gas. 
The hydrodynamic analysis in the superfluid phase can be found in Ref.~\cite{Hohenberg:1965}. 

\begin{figure}[tb]
\begin{center}
\includegraphics[width=8cm]{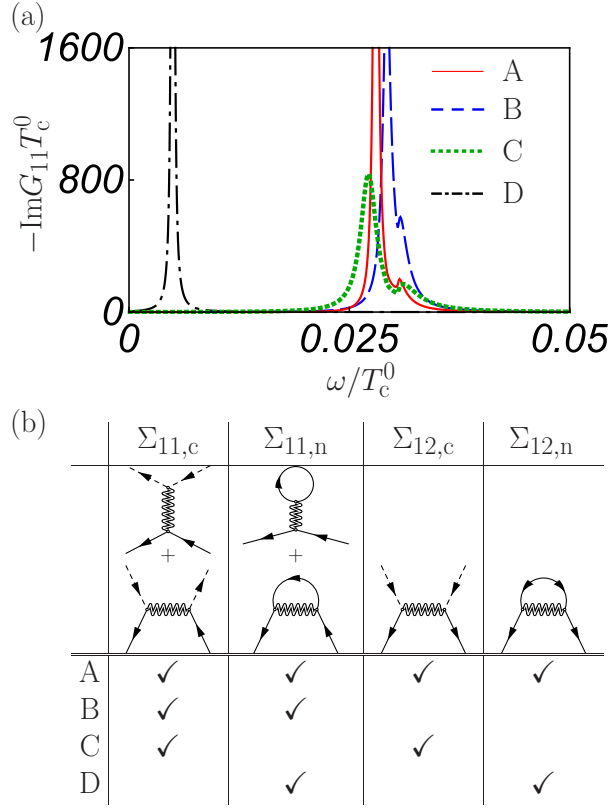} 
\end{center}
\caption{
Single-particle spectral function evaluated by various approximations.
(a) $G_{11}$ at $T = 0.5T_{\rm c}^{0}$ at $an^{1/3} = 10^{-2}$ and $q = 0.1q_{0}$. 
(b) Feynman diagrams used in panel (a). 
The solid arrow represents the single-particle line, the dashed arrow the condensate line, and 
the wiggly line the effective interaction line $U_{\rm eff}$ including the vertex correction in (\ref{RPABetheSalpeter}). 
}
\label{fig_graph5.fig}
\end{figure}

The origin of the satellite peak of the density response function can be discussed as follows: 
As shown in Figs.~\ref{fig_graph1.fig},~\ref{fig_graph2.fig},~\ref{fig_graph3.fig}, and ~\ref{fig_graph4.fig}, 
the satellite peak of the density response function $\chi$ is dominantly originated from the 1PR part $\chi^{\rm 1PR}$ that includes the single-particle Green's function through the density vertex $\Upsilon$. 
We thus separately treat the self-energy contribution in the single-particle Green's function to discuss the origin of the satellite peaks~\cite{Watabe:2018id}. 
The self-energy contribution in the BEC involves two-parts: diagonal and off-diagonal self-energies $\Sigma_{11(12)}$, 
which are also consists of two parts: condensate part $\Sigma_{11(12),{\rm c}}$ and non-condensate part $\Sigma_{11(12), {\rm n}}$. 
For $\Sigma_{12,{\rm n}}$, we consider the form $\Sigma_{12,{\rm n}} (p) = - \sum_{q} U_{\rm eff} (p) g_{12} (p+q)$. 
In order to separately analyze each contribution, we first consider the Hartree--Fock--Bogoliubov type self-energies, 
which can include all the contributions $\Sigma_{11(12),{\rm c}}$, and $\Sigma_{11(12),{\rm n}}$, diagrammatically described in Fig.~\ref{fig_graph5.fig}(b). 
In this approximation, the satellite peak can be seen (Fig.~\ref{fig_graph5.fig}(a)), which is consistent with the case of the Hartree--Fock--Bogoliubov--Popov type self-energies that does not include $\Sigma_{12,{\rm n}}$. 

The emergent satellite peak is possibly originated from (i) the off-diagonal self-energy $\Sigma_{12}$, (ii) non-condensate part $\Sigma_{11(12),{\rm n}}$, or (iii) condensate part $\Sigma_{11(12),{\rm c}}$. 
Figure~\ref{fig_graph5.fig}(a) shows the result of these contributions, where the self-energy contribution is selectively eliminated. 
The satellite peak still survives even if we eliminate the off-diagonal self-energy $\Sigma_{12}$, and the non-condensate part $\Sigma_{11(12),{\rm n}}$. 
On the contrary, the satellite peak vanishes when the condensate part of the self-energy $\Sigma_{11(12),{\rm c}}$ is absent. 
In the Bogoliubov approximation, where we replace the effective interaction $U_{\rm eff}(p)$ with the bare interaction $U$, the satellite and the broadening of the sharp peak structure never emerge, which gives the Bogoliubov excitation showing the sharp peak of the quasiparticle with infinite life-time. 
The origin of the satellite peak is thus concluded as the many-body BEC effect, namely, the condensate part of the self-energy, which gives the interaction between the condensate and the quasiparticle in the background of the many-body density-fluctuated medium. 
The non-condensate part of the self-energies, showing the quasiparticle-quasiparticle interaction effect, is not important for the satellite peak, where the many-body effect of the density fluctuation is smeared out by quasiparticles with various momenta.

\begin{figure}
\begin{center}
\includegraphics[width=8cm]{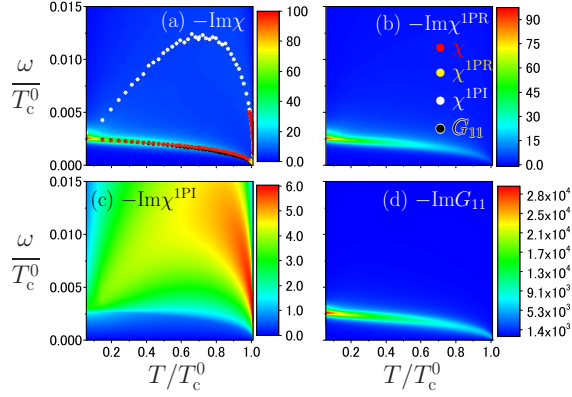} 
\end{center}
\caption{Frequency and temperature dependence of the response functions. 
(a) The total density response function $\chi$, (b) the 1PR part $\chi^{\rm 1PR}$, (c) the 1PI part $\chi^{\rm 1PI}$, and (d) the single-particle Green's function $G_{11}$. 
Red, yellow, white, and black points represent the maximum peak positions of $-{\rm Im}\chi$, $-{\rm Im}\chi^{\rm 1PR}$, $-{\rm Im} \chi^{\rm 1PI}$, and $-{\rm Im} G_{11}$, respectively. 
We used the same self-energies, density vertex, effective interaction and the 1PI part as shown in Fig.~\ref{fig_graph1.fig}. 
}
\label{fig_graph6.fig}
\end{figure}

One of the feature of the excitation of a BEC at $T=0$ is that 
the correspondence of the spectrum between the single-particle excitation and the collective excitation in the low-energy regime~\cite{Gavoret:1964gv}. 
We study the temperature dependence of these two excitations and discuss this correspondence by using the effective interaction including the vertex correction (Fig.~\ref{fig_graph6.fig}). 
Except close to the critical temperature, the density spectral function is dominated by the 1PR part, and thus the peak position of $\chi$ traces that of the 1PR part (Fig.~\ref{fig_graph6.fig}(a)). 
The intensity of the 1PI part is weak and its structure is very broad compared with the 1PR part (Figs.~\ref{fig_graph6.fig}(b) and (c)). 
The peak of the 1PI part is not monotonic function of the temperature, and close to $T_{\rm c}$, the peak of $\chi$ traces that of the 1PI part instead of the 1PR part, because the density vertex in the 1PR part becomes small. 
At the very low-temperature regime, we corroborated that the correspondence between the single-particle excitation peak and the collective excitation peak within the resolution of the numerical calculation. On the other hand, as the temperature increases, the peak of the single-particle excitation and that of the collective density excitation have a slight difference. This is due to the diminishing density vertex and the relatively increasing 1PI part as a function of the temperature.

\begin{figure}
\begin{center}
\includegraphics[width=8cm]{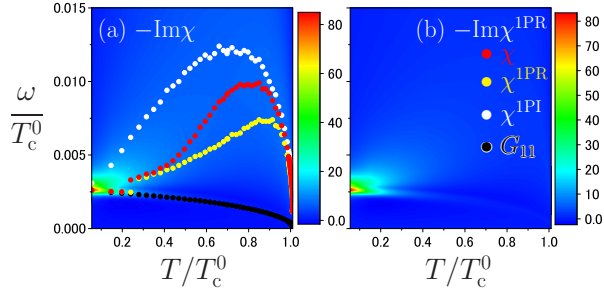} 
\end{center}
\caption{Temperature and frequency dependence of the response functions. 
(a) The total density response function $\chi$, and (b) the 1PR part $\chi^{\rm 1PR}$. 
Red, yellow, white, and black points are the same as in Fig.~\ref{fig_graph6.fig}. 
We used the same self-energies, effective interaction and the 1PI part as in Fig~\ref{fig_graph6.fig}.
For the density vertex, we employed $\Upsilon^{\rm s}$ and $\Upsilon^{{\rm s} \dag}$. 
}
\label{fig_graph7.fig}
\end{figure}

We discuss the approximation dependence on the result of the correspondence of the spectrum between the single-particle excitation and the density collective excitation (Fig.~\ref{fig_graph7.fig}). 
In contrast to the case of Fig.~\ref{fig_graph6.fig}, we employ the density vertex satisfying the identity $\Upsilon^{\rm s} (0) = 0$, 
where the vertex correction is omitted. 
In this case, the temperature dependence of the peak position of $\chi$ as well as $\chi^{\rm 1PR}$ 
are quite different from the case in Fig.~\ref{fig_graph6.fig}. 
As a result, the temperature dependence of the peak position may change, depending on approximations, such as the absence/existence of the vertex correction. 
However, in the very low temperature regime, we can still find the correspondence between the peak positions between the density response function and the single-particle Green's function.

\begin{figure}
\begin{center}
\includegraphics[width=10cm]{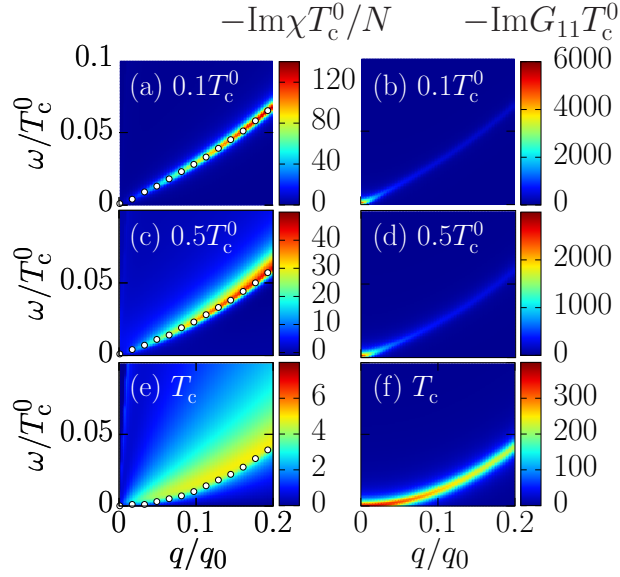} 
\end{center}
\caption{Momentum and frequency dependence of the total density response function $\chi$, and the single-particle Green's function $G_{11}$. 
White points in panels (a), (c), and (e) represent the maximum peak position of $-{\rm Im} G_{11}$. 
We used the self-energies in (\ref{eq27s}) and (\ref{eq28s}), with the effective interaction (\ref{RPABetheSalpeter}) including the vertex correction. 
For the density vertex, we employed $\Upsilon^{\rm s}$ and $\Upsilon^{{\rm s} \dag}$. 
We used the 1PI part in (\ref{Eq259}) including the vertex correction.  
}
\label{fig_graph8.fig}
\end{figure}

The density response function and the single-particle spectral function are shown in the $\omega$-$q$ plane in Fig.~\ref{fig_graph8.fig}. 
The peak of the single-particle excitation traces that of the density response function at low temperature ($0.1T_{\rm c}^{0}$). 
This correspondence cannot be seen at $T_{\rm c}$, because of the absence of the BEC. 
At moderate temperature ($0.5T_{\rm c}^{0}$), although two peak positions are slightly different at high-momentum and high-energy regime, the correspondence may survives in the low-momentum and low-energy regime. 
At very low temperature, the peak position is well described by the Bogoliubov approximation, although the satellite peak emerges which is not reproduced by the mean-field type Bogoliubov approximation~\cite{Watabe:2018id}. 
As temperature increases, the width of the single-particle spectral function becomes broad, and the phonon structure disappears at $T_{\rm c}$. 
The density response function at higher temperature also becomes quite broad. 
From these results, we can reasonably expect that it is an essential feature in BECs that the peak position of the density response function overlaps with that of the single-particle Green's function not only at the zero temperature but also in the very low but nonzero temperature regime, which is irrespective of the approximation that we take. 
The linear dispersion at $T=0$ is analytically discussed to be originated from the identity $\partial_{\omega} \Sigma_{11} (0) = 1$~\cite{Nepomnyashchii:1978wb} for the theory satisfying $\Sigma_{12} (0) = 0$. 
In many-body approximations at nonzero temperatures, 
the numerical analytic continuation makes it difficult to analyze the origin of the structure of the excitation spectrum. 
Although the linear dispersion can be originated from $\Sigma_{12} (0)$ in the approximation $\Sigma_{12} (0) \neq 0$ as discussed in Ref.~\cite{abrikosov1975methods}, this problem is important all the more in the many-body approximation at nonzero temperatures satisfying the identity $\Sigma_{12} (0) = 0$~\cite{Watabe:2018id}. 

The sound speed can be estimated by inversely solving the compressibility zero-frequency sum-rule $c = \sqrt{- n /[m \chi^{} (0)]}$. 
Since this sum-rule is exhausted by the 1PI part because of $\chi^{\rm 1PR} (0) = 0$, 
the sound speed is exactly given by \begin{align}
c = \sqrt{- \frac{n}{m} \frac{1}{\chi^{\rm 1PI} (0)}} . 
\label{eq333}
\end{align} 
If we employ $\Upsilon^{\rm s}$ and $\Upsilon^{{\rm s} \dag}$, we can reproduce the exact identity $\chi^{\rm 1PR} (0) = 0$, because of $\Upsilon^{\rm s} (0) = 0$. 
The 1PI part (\ref{Eq259}) in the static and low-momentum limits is given by 
\begin{align}
\chi^{\rm 1PI} (0) = 
- \frac{1}{U} \frac{ 1 - U \Pi' (0) }{3 - 2 U \Pi' (0)}, 
\end{align}  
and the sound speed at $T \leq T_{\rm c}$ can be estimated as 
\begin{align}
c = c_0 \sqrt{
\frac{3 - 2 U \Pi' (0) }
{ 1 - U \Pi' (0)  } 
}, 
\label{Eq306}
\end{align} 
where $c_0 \equiv \sqrt{Un/m}$. 
This sound speed (\ref{Eq306}) is found to be a positive real number if we are considering the repulsive interaction $U > 0$, because $\Pi' (0)$ given in (\ref{eq88Pi}) is a real negative number according to the relation $\partial n_{\bf p} / \partial E_{\bf p} = - \beta n_{\bf p} (1 + n_{\bf p}) < 0$. 
At $T \geq T_{\rm c}$, the sound speed is given by 
\begin{align}
c = c_0 \sqrt{  \frac{1 - 2 U \Pi_{22} (0) }{ - U \Pi_{22} (0) }  }, 
\label{H10}
\end{align}
where the 1PI part is given in (\ref{eq98}). 
This sound speed (\ref{H10}) is also safely a positive real number for $U>0$, because of the relation 
\begin{align}
 \lim\limits_{{\bf q} \rightarrow {\bf 0}}\Pi_{22} (0,{\bf q}) 
= & 
 \sum\limits_{\bf p} 
\frac{\partial n_{\bf p}' }{ \partial \varepsilon_{\bf p}} 
= 
- 
\beta \sum\limits_{\bf p} 
n_{\bf p}' (1 + n_{\bf p}') < 0. 
\end{align} 

One may employ the simpler regular part (\ref{Eq259simpler}) for the 1PI part (\ref{Eq259}). 
In this bubble diagram case not including the vertex correction, however, we obtain an unphysical temperature-independent sound speed $c = c_0 = \sqrt{Un/m}$ for all temperatures below $T_{\rm c}$. 
Since $\chi_{\rm R}^{\rm s} (0)$ exhibits the infrared divergence at $T \leq T_{\rm c}$, 
the 1PI part (\ref{Eq259}) in the static and low-momentum limits is temperature-independent, given by $\chi_{\rm s}^{\rm 1PI} (0) = - 1/U$.

\begin{figure}[tb]
\centering
\includegraphics[width=8cm]{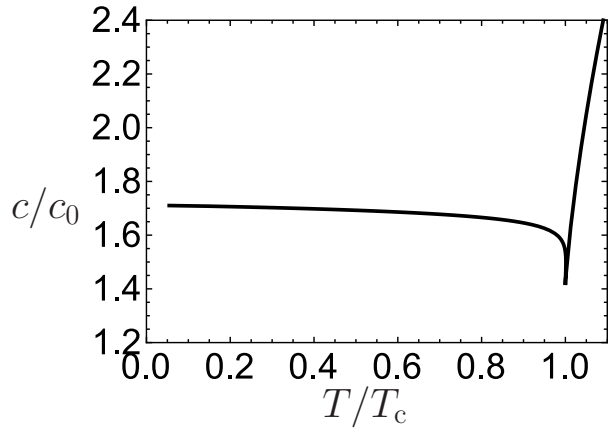} 
\caption{Sound speed evaluated from the zero-frequency compressibility sum-rule $c = \sqrt{-n / [m \chi^{\rm 1PI} (0)]}$, scaled by $c_0 = \sqrt{U n/m}$. We used the relations (\ref{Eq306}) and (\ref{H10}) below and above the critical temperature, respectively. }
    \label{fig:FigSound}
\end{figure}

We discuss the temperature dependence of the sound speed $c$ using the RPA  (\ref{Eq306}) and (\ref{H10}) (See Fig.~\ref{fig:FigSound}). 
In this formalism, the sound speed is temperature dependent, and the sound speeds in (\ref{Eq306}) and (\ref{H10}) merge at $T = T_{\rm c}$, because of the infrared divergence of the correlation functions $\Pi' (0)$ and $\Pi_{22} (0)$ at this temperature.  
The sound speed is given by $c  = \sqrt{2}c_0$ at $T = T_{\rm c}$ within the RPA including the vertex correction, where 
the factor $2$ comes from the many-body effect in this approximation. 
In the Bogoliubov-Popov mean-field calculation, the sound speed is given by $c = \sqrt{U n_0/m}$, and it drops to zero at the critical temperature. 
At absolute zero temperature case, the sound speed (\ref{Eq306}) is approximately given by $c  = \sqrt{3}c_0$ for $p_{\rm c} a \ll 1$. 
The sound speed is overestimated in the RPA with the vertex correction, although it reproduces the same order of the sound speed in the Bogoliubov approximation at $T =0$. For the consistency, further improvements may be necessary for the calculation of the sound speed derived from the RPA with the zero-frequency compressibility sum-rule. 

The sound velocity of the liquid $^4$He has been experimentally studied above the critical temperature~\cite{Findlay1938,Atkins1951,Chase1953,Itterbeek1954} and below the critical temperature~\cite{Findlay1938,Pellam1947,Atkins1951,Chase1953,Itterbeek1954,Chase1958,Whitney1962,Chase1964,Whitney1967,Abraham1969,Pike1970,Winterling1973,Eselson1974,Maza1988}. The measurement of the attenuation is also reported~\cite{Pellam1947,Atkins1959,Jeffers1965,Abraham1969,Winterling1973,Eselson1974}. 
Above the critical temperature, the temperature dependence of the sound velocity is convex~\cite{Findlay1938,Atkins1951,Itterbeek1954}. 
On the other hand, below the critical temperature, the sound velocity is slightly increased for increasing temperature and decreases rapidly near the $\lambda$-point~\cite{Chase1958,Atkins1959,Whitney1962,Whitney1967,Abraham1969,Eselson1974}. The maximum value of the sound velocity is measured around 0.7K~\cite{Whitney1962,Whitney1967}. 
At the critical temperature, the sound velocity shows a cusp anomaly~\cite{Findlay1938,Atkins1951,Itterbeek1954,Whitney1962,Whitney1967,Abraham1969,Pike1970,Eselson1974,Maza1988}. 

There has been a debate whether the sound speeds below and above the critical temperature converge to the same value or show the discontinuity at the critical temperature in superfluid $^4$He. 
The measurement of the sound velocity very close to the $\lambda$-point has the fundamental difficulty~\cite{Atkins1951, Chase1958,Atkins1959}. 
No detectable discontinuity of the sound velocity was discussed at the $\lambda$-transition~\cite{Findlay1938,Atkins1951,Atkins1959}. 
The specific heat shows the jump, which suggests the second order phase transition according to the Ehrenfest relations~\cite{Atkins1959}, 
and the isothermal compressibility $\kappa_T$ shows not the divergence but a discontinuity at the transition point~\cite{Lounasmaa1963}. 
On the other hand, the logarithmic singularity of $\kappa_T$ is also discussed at the $\lambda$-transition~\cite{Chase1964,Simanta2006}. 
The sound velocity near the $\lambda$-point is also theoretically investigated~\cite{Ferrell1980,Ferrell1982}, and ultrasonic attenuation is also studied based on the Pippard--Buckingham--Fairbanks relations~\cite{Ferrell1980}. 
Within the present formalism, the sound speeds converge to the same value from above and below the critical temperature, 
where it should be noted that thoughtful treatments are needed in fluctuation regions~\cite{CapogrossoSansone:2010kta}. 

In the formalism used in this paper, we take the Hartree--Fock--Bogoliubov--Popov approximation $g(p)$ for constructing the building blocks and self-energies. 
One of the directions for the future study is to develop the self-consistent approximation, such as the self-consistent $T$-matrix approximation~\cite{Haussmann:2007ia}. 
In contrast to the Fermi gas, the BEC provides the infrared divergence in the single-particle Green's function with a relation $G_{11} (0) = - G_{12}(0)$, which also provides a strong constraint for the infrared divergent polarization functions, given by $\Pi_{11,22,14} (0) = - \Pi_{12} (0)$~\cite{Watabe:2013hw}. 
Since the exact infrared property is important for studying the low-energy properties of the BEC~\cite{Watabe:2014gv}, this constraint will be important in development of the self-consistent approximation for the BEC.

The matrix formalism for BECs presented in this paper will have potential to efficiently study the exact low-energy properties of the single-particle Green's function and the density response function at nonzero temperatures as an extension of the theory at $T=0$ by Gavoret and Nozi\`eres~\cite{Gavoret:1964gv}. 
In this nonzero temperature case, we will need the forth order expansion of the self-energy with respect to $\omega$ and $p_i$, in order to study the second sound contribution~\cite{Hohenberg:1965}. 
Even in this case, the Bogoliubov operator inequality and the Josephson sum-rule are still important criteria for checking the validity of the results. 
The present matrix formalism will also have potential to extend theories for the spinor BEC~\cite{Kawaguchi:2012bl}, the dipolar BEC~\cite{Baranov:2008cw}, the collisionless sound~\cite{Ota:2018fc}, the deep inelastic scattering~\cite{Hofmann2017}, the Bose polaron problem~\cite{CamachoGuardian:2018hl}, and the renormalization-group method~\cite{Bijlsma1996,Stoof:2013dva,kopietz2010introduction}. 

Ultracold atomic gases may serve as a platform for directly addressing the strong connection between the single-particle and density excitations in BECs by employing useful tools, such as the Feshbach resonance, the uniform box trap, and the spectroscopy. Theoretical concepts of BECs that should be interesting to confirm experimentally are the Josephson sum-rule, as well as the equivalence of the dispersion relations between the single-particle and collective excitations. 
It is also interesting to experimentally study the phonon-maxon-roton excitation in dipolar BECs not only in the collective excitation~\cite{Petter:2019fx}, but also in the single-particle excitation below and above the critical temperature by controlling the relative strength of the dipolar to the contact interactions~\cite{Petter:2019fx}. 
Since the sharp maxon-roton intensity has been considered to originate from the single-particle excitation and the BEC in the superfluid $^{4}$He~\cite{Glyde1990}, it will provide deeper understanding of the maxon-roton excitations as well as the connection between the single-particle and density excitations in BECs, with extending the context of superfluid $^4$He.

\section{Conclusions}\label{SecVIII}

We investigated the single-particle excitation and the collective density excitation in Bose--Einstein condensates (BECs) by using the single-particle Green's function and the density response function. 
First, we revisited the earlier study presented by Gavoret and Nozi\`eres~\cite{Gavoret:1964gv}, with including the subsequent results given by Nepomnyashchii and Nepomnyashchii~\cite{Nepomnyashchii:1975vs,Nepomnyashchii:1978wb}. 
We extended the Nambu representation of the single-particle Green's function for BECs to correlation functions and vertex functions by making the use of the matrix formalism, which reproduces the exact properties efficiently. 
By following the discussion given by Gavoret and Nozi\`eres~\cite{Gavoret:1964gv} with the matrix formalism, we revisited the low-energy properties of the correlation functions and the vertex functions, and the correspondence of the spectrum between the single-particle excitation and the collective excitation in the low-energy and low-momentum regime. 
We also present an overview of the earlier experimental and theoretical studies on the collective excitations in superfluid $^4$He as well as in ultracold atomic gases. We also gave criticisms on theories casting doubt upon the conventional wisdom of the BEC: the equivalence of the dispersion relations between the single-particle excitation and the collective excitation in the low-energy and low-momentum regime. 
The consistency with the Bogoliubov operator inequality and the Josephson sum-rule is an important criterion for the theory contradict to the conventional wisdom.

By applying the matrix formalism, we developed a random phase approximation (RPA) for BECs to describe a single-particle Green's function and the density response function at nonzero temperatures. 
Depending on the presence or absence of the vertex correction, 
approximations provide the quantitatively different temperature dependence of the density response function and the single-particle spectral function. 
However, the peak positions in both functions are consistent in the very low-temperature regime, 
which supports the correspondence of the spectrum between the single-particle excitation and the collective excitation. 
Many-body effect can be seen in the satellite structure of the single-particle spectral function, which comes from the interaction between the condensate and the quasiparticles in the medium with the density fluctuation. 
By using the the compressibility zero-frequency sum-rule, the temperature dependence of the sound speed was evaluated, 
where the result within the RPA including the vertex correction shows no discontinuity at the critical temperature, although careful treatments are necessary in the fluctuation region.

\acknowledgments

Useful conversations with Y. Kato, Y. Ohashi, M. Ueda, and T. Nikuni are acknowledged in the very initial stage of this study. 
The author thanks A. J. Leggett for providing information about studies by T. Kebukawa and by R. A. Ferrell. 
The author was supported by JSPS KAKENHI Grant No. 249416, JP16K17774, JP18K03499. 

\appendix

\section{Thermodynamic relations}\label{AppendixB}

We summarize thermodynamic relations with the use of the Nepomnyashchii--Nepomnyashchii identity (\ref{eq52}). 
We also show relations with respect to the isothermal sound speed $c_{\rm T}$. 

Since the thermodynamic potential $\Omega' = - T \ln \Xi$ is related to the grand potential $\Omega$ through $\Omega' = \Omega + \mu_0 n_0$, we have a relation $d \Omega' (T, \mu, n_{0}) =  - S dT - n' d \mu + \mu_{0} dn_{0}$, 
where the volume of the system is assumed to be fixed. 
Here, the entropy $S$, the non-condensate density $n'$ and the chemical potential of the condensate $\mu_{0}$, which satisfies $\mu = \mu_0$, are respectively given by 
$S = - \partial \Omega' / \partial T$, 
$n' = - \partial \Omega' / \partial \mu$, and 
$\mu_{0} = \partial \Omega' / \partial n_{0}$.  

A thermodynamic relation provides 
\begin{align}
& d \mu_{0} (T, \mu,n_0) 
\nonumber \\
= & 
\left .  \frac{\partial^{} \mu_{0}}{\partial T} \right |_{\mu,n_0} d T 
+ 
 \left . \frac{\partial^{} \mu_{0}}{\partial \mu} \right |_{T,n_0} d \mu 
+ 
\left . \frac{\partial^{} \mu_{0}}{\partial n_0} \right |_{T,\mu} d n_0. 
\label{eqB3}
\end{align}
By using the relations $d\mu_0 = d\mu$ and $\partial^2 \Omega' / (\partial T \partial n_0) = \partial \mu_0 / \partial T |_{\mu, n_0} = - \partial S / \partial n_0 |_{\mu, n_0}$, as well as  
the Nepomnyashchii--Nepomnyashchii identity (\ref{eq52}), we obtain identities 
\begin{align}
\frac{\partial^{2} \Omega'}{\partial\mu \partial n_{0}} = \left . \frac{\partial \mu_0}{\partial \mu} \right |_{T,n_0} = - \left . \frac{\partial n'}{\partial n_{0}} \right |_{T,\mu} = 1 + \left . \frac{\partial S}{\partial n_0} \right |_{T,\mu} \frac{d T}{d \mu}. 
\label{eqB4}
\end{align} 
Given (\ref{eq47}) as well as (\ref{eqB4}), we have 
\begin{align}
\Upsilon_{0} (0) = & - \sqrt{n_{0}} 
\left . \frac{\partial S}{\partial n_0}  \right |_{T,\mu} \frac{d T}{d \mu}
| + \rangle ,  
\end{align}
In the isothermal condition, we end with $\Upsilon_0 (0) = 0$ as shown in (\ref{eq54}).

We also have other thermodynamic relation 
\begin{align}
d n(T, \mu,n_0) = 
\left . \frac{\partial n}{\partial T} \right |_{\mu, n_0} d T
+ 
\left . \frac{\partial n}{\partial \mu} \right |_{T,n_0} d \mu 
+ 
\left . \frac{\partial n}{\partial n_0} \right |_{T,\mu} d n_0. 
\label{eqB6}
\end{align}
Since $n_0$ is fixed in the first and second terms in (\ref{eqB6}), 
we have relations 
$\partial n / \partial T |_{\mu, n_0} = \partial n' / \partial T |_{\mu, n_0} = \partial S / \partial \mu |_{T, n_0}$ and 
$ \partial n/\partial \mu |_{n_0} = \partial  n'/\partial \mu |_{n_0} $. 
Since $n = n_0 + n'$, we also have 
\begin{align}
\left . \frac{\partial n}{\partial n_0} \right |_{T,\mu}= 
1 + \left . \frac{\partial n'}{\partial n_0} \right |_{T,\mu} 
= - \left . \frac{\partial S}{\partial n_0} \right |_{T, \mu} \frac{dT}{d \mu}, 
\label{eqB7}
\end{align}
where we applied (\ref{eqB4}) to the last equality. 
The thermodynamic relation is then reduced into 
\begin{align}
\frac{d n}{d\mu} = 
\left . \frac{\partial n'}{\partial \mu} \right |_{T, n_0}
+ 
\left ( 
\left . \frac{\partial S}{\partial \mu} \right |_{T,n_0}
- 
\left . \frac{\partial S}{\partial n_0} \right |_{T, \mu} \frac{d n_0}{d \mu} 
\right ) 
\frac{dT}{d \mu} . 
\label{eqB7a}
\end{align} 
As a result, we have a thermodynamic relation with respect to the isothermal sound speed $c_{\rm T}$, giving the form 
\begin{align} 
\frac{n}{mc_{\rm T}^{2}} = \left . \frac{d n }{d\mu} \right |_{T} = 
\left . \frac{\partial n'}{\partial \mu} \right |_{T,n_{0}} = - \frac{\partial^2 \Omega}{\partial \mu^2}, 
\label{eqB5}
\end{align} 
where the second equality is obtained from \eqref{eqB7a} with the isothermal condition.

\section{Low energy behaviors of single-particle Green's function}\label{AppendixBC}

The Dyson equation (\ref{eq11}) provides the single-particle Green's function, given in the form 
\begin{align}
G (p) 
= & \frac{1}{D(p)} 
[\omega \sigma_{3} +\varepsilon_{\bf p} -\mu+ \sigma_{3} \Sigma (-p) \sigma_{3}] , 
\label{eq71}
\end{align}
where $D (p) \equiv [ D_{0}^{2} (p) - D_{+}(p) D_{-}(p) ] / 4$, 
with $D_{0} (p) \equiv {\rm Tr}   [  \omega -  \sigma_{3} \Sigma (p)]$ and $D_{\pm} (p) \equiv \langle \pm | [ \varepsilon_{\bf p} - \mu + \Sigma (p) ] | \pm \rangle$. 
In the low energy regime, by using (\ref{eq66}), (\ref{eq68}) and (\ref{eq69}), 
we obtain $D_{0} (p) = O (p^{2})$ and 
\begin{align}
D_{-} (p) \simeq & 
\frac{1}{2} \langle - | \partial_\omega^2 \Sigma ' (0) | - \rangle \omega^2 
+ 
\left [ \frac{1}{m} + \frac{1}{2}  \langle - | \partial_{\bf p}^2 \Sigma ' (0) | - \rangle \right ] {\bf p} ^2 
\nonumber
\\ 
= &  - \frac{n}{n_0 mc_{\rm T}^2} (\omega^2 - c_{\rm T}^2 {\bf p}^2) . 
\label{eq77}
\end{align} 
By using the fact that the leading term of the off-diagonal self-energy $\Sigma_{12}$ is the nonanalytic part $\Delta \Sigma (p)$ in the small $p$ regime, we also have 
$D_{+} (p) \simeq 4 \Delta \Sigma_{} (p) \simeq 4 \Sigma_{12} (p)$. 
As a consequence, relations in the low energy regime 
\begin{align}
\begin{pmatrix} 
G_{11} \pm G_{12} 
\\
G_{21} \pm G_{22} 
\end{pmatrix} =
G(p) | \pm \rangle 
\simeq & - \frac{2}{D_\pm (p) } | \pm \rangle 
\label{eq78}
\end{align} 
provides $G_{11,22} = -1/D_- - 1/D_+$ as well as $G_{12,21} = +1/D_- - 1/D_+$. 
We thus end with Eq. \eqref{eq70}. 

\section{Derivations of (\ref{eq92}) and (\ref{eq84})-(\ref{eq87}) }\label{AppendixC}

\subsection{derivations of (\ref{eq92})} 

An element of the density and current vertices are given by 
$
\gamma_{\mu} (q) 
=  
[ \Upsilon_{\mu}^{\dag} (q) | 0 \rangle 
+ 
\Upsilon_{\mu}^{\dag} (-q) | 1 \rangle
]/2$, 
where we used the symmetry relations of the elements in $\Upsilon_{\mu}^{\dag} (q)$. 
Using ({\ref{eq41}), we find that 
\begin{align}
\gamma_{\mu} (q) 
= & - \frac{1}{2} T 
\sum\limits_{p}  \langle \lambda_{\mu} (p;q) | Q (p;q) | 0 \rangle 
\nonumber \\ & 
- \frac{1}{2} T 
\sum\limits_{p}  \langle \lambda_{\mu} (p;-q) | Q (p;-q) | 1 \rangle . 
\label{eqC10}
\end{align}
Given (\ref{eq25}) as well as (\ref{eq55}), this vertex is constructed from three parts: 
$\gamma_{\mu} (q) = \gamma_{\mu}^{(1)} (q) + \gamma_{\mu}^{(2)} (q) + \gamma_{\mu}^{(3)} (q)$, 
where 
\begin{align}
\gamma_{\mu}^{(1)} (q) 
= &  
- \frac{1}{4} \sqrt{-1} 
[ 
\langle \lambda_{\mu} (0;q) | {\mathcal G}_{1/2} | 0 \rangle 
\nonumber \\ & 
+ \langle \lambda_{\mu} (0;-q) | {\mathcal G}_{1/2} | 1 \rangle 
]  , 
\label{eqC13}
\\ 
\gamma_{\mu}^{(2)} (q) 
= &  
- \frac{1}{4} \sqrt{-1} 
[ 
 \langle \lambda_{\mu} (-q;q) | \hat X {\mathcal G}_{1/2} | 0 \rangle 
 \nonumber \\ & 
+ 
\langle \lambda_{\mu} (q;-q) | \hat X {\mathcal G}_{1/2} | 1 \rangle 
]  , 
\label{eqC14}
\\ 
\gamma_{\mu}^{(3)} (q) 
= & 
- \frac{1}{4} T  \sum\limits_{p} 
[ 
\langle \lambda_{\mu} (p;q)  | L (p;q) | 0 \rangle 
\nonumber \\ & 
+ 
\langle \lambda_{\mu} (p;-q)  | L (p;-q) | 1  \rangle 
] . 
\label{eqC12}
\end{align}

The terms $\gamma_{\mu}^{(1)} (q) $ and $\gamma_{\mu}^{(2)} (q)$ are reduced to 
$\gamma_{\mu}^{(1)} (q) = \gamma_{\mu}^{(2)} (q) =  \lambda_{\mu} (0;q) \sqrt{n_{0}}^{} /2$, 
where we used relations 
$\lambda_{\mu} (0;-q) = f_{\mu} \lambda_{\mu} (0;q)$, $\lambda_{\mu} (\mp q;\pm q) = \lambda_{\mu} (0;\mp q)$ as well as 
$
\langle f_\mu | {\mathcal G}_{1/2} | 0 \rangle = 
\langle f_\mu | \hat X {\mathcal G}_{1/2} | 1 \rangle  
=  
f_{\mu} \langle f_\mu | {\mathcal G}_{1/2} | 1 \rangle = 
f_{\mu} \langle f_\mu | \hat X {\mathcal G}_{1/2} | 0 \rangle 
= 
\sqrt{-n_{0}} . 
$ 
The sum of these two terms $\gamma_{\mu}^{(1)} + \gamma_{\mu}^{(2)}$ provides the first term of (\ref{eq92}).

The term $\gamma_{\mu}^{(3)} (q) $ in the first order of $q$ is given by 
\begin{align}
\gamma_{\mu}^{(3)} (q) 
\simeq & 
- \frac{T}{4 } \sum\limits_{p} 
\lambda_{\mu} (p;0)  
\left [ 
\langle f_\mu | L (p;q) | 0 \rangle 
+ 
\langle f_\mu | L (p;-q) | 1 \rangle  
\right ] , 
\label{eqC15}
\end{align} 
where we have used $\langle f_\mu | L (p;0) | 0 \rangle = \langle f_\mu | L (p;0) | 1 \rangle$ as well as 
$\lambda_{i} (p;\pm q) = \lambda_{i} (p;0) \pm \lambda_{i} (0;q)$ for $i = 1,2,3$. 
Using (\ref{eq57}), we find 
\begin{align}
\gamma_{\mu}^{(3)} (q) 
\simeq & 
- \frac{T}{4 } \sum\limits_{p} 
\lambda_{\mu} (p;0) 
\nonumber \\ & \times 
[ 
{\mathcal D}_1 (q) \langle f_\mu | {\bf G} (p) 
+ 
{\mathcal D}_2 (q) \langle f_\mu | \hat B {\bf G} (p)  
]. 
\label{eqC16}
\end{align} 
By using (\ref{eqC16}) as well as the following two mathematical identities 
$\langle f_{\mu } | {\bf G}(p) = {\rm Tr} [\sigma_{\mu}' G (p) ]$, 
and $\langle f_{\mu } | \hat B {\bf G}(p) = - {\rm Tr} [\sigma_{\mu}' \sigma_{3} G (p) ]$, 
we obtain the second term of (\ref{eq92}). 
We can thus obtain (\ref{eq92}).  

We can also derive the same result by using 
$\gamma_{\mu} (q) 
=  
[ \langle 0 | \Upsilon_{\mu} (q) 
+ 
\langle 1 | \Upsilon_{\mu} (-q)  
]/2$.  
In this case, we apply a variant of (\ref{eq57}), giving the form 
$\langle 0 | L^\dag (p, + q)  + \langle 1 | L^\dag (p, - q) 
\simeq  
\hat {\mathcal D} (q) 
{\bf G}^\dag (p) $,  
where $L^\dag (p; q) = P^\dag (p;q) K_0 (p;q)$. 
We also apply the mathematical identities 
$ {\bf G}^\dag (p) | f_\mu \rangle = {\rm Tr} [\sigma_{\mu}' G (p) ]$, 
and 
${\bf G}^\dag (p) \hat B | f_\mu \rangle  = - {\rm Tr} [\sigma_{\mu}' \sigma_{3} G (p) ]$, 
as well as 
$
\langle 0  | {\mathcal G}_{1/2}^\dag | f_\mu \rangle = 
\langle 1 | {\mathcal G}_{1/2}^\dag  \hat X | f_\mu \rangle  
=  
f_{\mu} \langle 1 | {\mathcal G}_{1/2}^\dag | f_\mu \rangle = 
f_{\mu} \langle 0 | {\mathcal G}_{1/2} ^\dag  \hat X | f_\mu \rangle 
= 
\sqrt{-n_{0}}
$.

\subsection{derivations of (\ref{eq84})-(\ref{eq87})} 

We derive the low energy behavior of the 1PI part $\chi_{\mu\nu}^{\rm 1PI}$. 
First, we can reduce Eq. (\ref{eq37new}) into 
\begin{align} 
\chi_{\mu\nu}^{\rm 1PI} (q) 
= & 
- \frac{1}{2 } T
\sum\limits_{p} 
\langle  \lambda_{\mu} (p;q) | K_{0} (p;q) | \Lambda_{\nu} (p;q) \rangle,
\label{eq37} 
\end{align} 
where we introduced the density and current vertex vector with the vertex corrections, given by 
\begin{align}
| \Lambda_{\nu} (p;q)  \rangle= & 
| \lambda_{\nu} (p;q) \rangle 
- 
T \sum\limits_{p'} 
\Gamma (p,p';q) K_{0} (p';q) | \lambda_{\nu} (p';q) \rangle. 
\label{eq39} 
\end{align} 
The four point vertex $\Gamma$ can be related to the two point vertex ${\bf \Sigma}$, where two of four vertex points are blocked by the single-particle Green's function ${\bf G}$. 
By taking the derivative $\delta / \delta x_\nu$, we have~\cite{Gavoret:1964gv} 
\begin{align}
\frac{\delta }{\delta x_{\nu}} 
{\bf \Sigma} (p)  
= 
& 
f_\nu T
\sum\limits_{p'} 
\Gamma (p,p';0) K_{0} (p'; 0) |\lambda_{\nu} (p'; 0)  \rangle, 
\label{eqC2}
\end{align} 
which is diagrammatically described in Fig.~\ref{fig12.fig}. 
The factor $f_{\nu} \lambda (p;0)$ as well as the bare part of the two-particle Green's function $K_0$ come from a relation $\delta G_{0} (p) /\delta x_{\nu} = - f_{\nu} \lambda (p;0) G_{0}^{2} (p)$~\cite{Gavoret:1964gv}. 
As a result, the density/current vertex vector with the vertex corrections at $q = 0$ is reduced into 
$| \Lambda_{\nu} (p;0)  \rangle = 
| \lambda_{\nu} (p;0)  \rangle - 
f_{\nu} \delta {\bf \Sigma} (p) / {\delta x_{\nu}}$. 

\begin{figure}[t]
\begin{center}
\includegraphics[width=6cm]{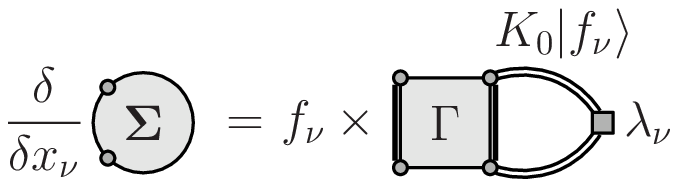} 
\end{center}
\caption{
Diagrammatic representation of (\ref{eqC2}). 
}
\label{fig12.fig}
\end{figure}

The 1PI part $\chi_{\mu\nu}^{\rm 1PI}$ is then given in the form 
\begin{align}
\chi_{\mu\nu}^{\rm 1PI} (0) 
= & 
- \frac{1}{2 } T
\sum\limits_{p} 
\langle  \lambda_{\mu} (p;0) | K_{0} (p;0) 
| \lambda_{\nu} (p;0)  \rangle 
\nonumber
\\
&
+ 
\frac{1}{2 } T
\sum\limits_{p} 
\langle  \lambda_{\mu} (p;0) | K_{0} (p;0) 
f_{\nu} \frac{\delta}{\delta x_{\nu}} {\bf \Sigma} (p) . 
\label{eqC3}
\end{align} 
We may also have two mathematical identities 
\begin{align}
&  
\langle \lambda_{\mu} (p; 0) | K_{0} (p;0) | \lambda_{\nu} (p;0) \rangle 
\nonumber 
\\ 
= & 
\lambda_{\mu} (p; 0) \lambda_{\nu} (p;0) 
{\rm Tr} \left [ \sigma_{\mu}' G (p) \sigma_{\nu}' G (p) \right ] , 
\label{eqC4}
\\ 
&
\langle f_{\mu} | 
K_{0} (p;0) \frac{\delta }{\delta x_{\nu}} {\bf \Sigma} (p) 
= 
{\rm Tr} \left [ 
\sigma_{\mu}'
G (p) \frac{\delta \Sigma (p)}{\delta x_{\nu}} G (p)
\right ]. 
\label{eqC5}
\end{align} 
Given these identities, we may reduce $\chi_{\mu\nu}^{\rm 1PI} (0) $ into 
\begin{align}
\chi_{\mu\nu}^{\rm 1PI} (0) 
= & 
- \frac{1}{2 } T
\sum\limits_{p} 
{\rm Tr} \biggl \{
f_\nu 
 \lambda_{\mu} (p; 0) \sigma_{\mu}' G (p) 
\nonumber \\ & \times 
\biggl [
f_\nu \lambda_{\nu} (p;0) 
\sigma_{\nu}' 
- 
 \frac{\delta \Sigma (p)}{\delta x_{\nu}}
\biggr ]
G (p) 
\biggr \} . 
\label{eqC6}
\end{align}

From the Dyson-Beliaev equation $G = G_{0} + G_{0} \Sigma G$, we can derive~\cite{Gavoret:1964gv} 
\begin{align}
\frac{\delta G (p) }{\delta x_{\nu}} 
= & 
- 
G (p)
\left [ 
\frac{\delta G_0^{-1} (p) }{\delta x_{\nu}} 
- 
\frac{\delta { \Sigma} (p) }{\delta x_{\nu}} 
\right ] { G} (p) . 
\label{eqC7}
\end{align} 
In particular, we have 
$\delta G_0^{-1} (p) / \delta x_{\nu} = f_\nu \lambda_{\nu} (p;0) \sigma_{\nu} '$. 
By applying these relations to (\ref{eqC6}), we end with 
\begin{align}
\chi_{\mu\nu}^{\rm 1PI} (0) 
= & 
\frac{1}{2} T 
\sum\limits_{p}  f_{\nu} \lambda_{\mu} (p;0) 
\frac{\delta}{\delta x_{\nu}}  {\rm Tr} [ \sigma_{\mu} ' G(p) ] , 
\label{eq82}
\end{align}
where $\sigma_{\mu} ' = {\rm diag} (1, f_{\mu} )$. 
This relation provides \eqref{eq84}, \eqref{eq85}, \eqref{eq86}, and \eqref{eq87}.

\section{Polarization Functions}\label{AppendixD} 
\par 
We summarize the polarization functions for the random-phase approximation studied in this paper~\cite{Watabe:2013hw,Watabe:2014gv,Watabe:2019fz}. 
At $T \leq T_{\rm c}$, the polarization functions are given by 
\begin{widetext}
\begin{align} 
\Pi_{11} (q) = & 
- 
\sum\limits_{\bf p} 
\frac{1}{2} \left [ (E_{{\bf p}+{\bf q}} - E_{\bf p}) \left ( 1 - \frac{\xi_{{\bf p}+{\bf q}}\xi_{\bf p}}{E_{{\bf p}+{\bf q}}E_{\bf p}} \right ) + i \omega_{n} \left ( \frac{\xi_{{\bf p}+{\bf q}}}{E_{{\bf p}+{\bf q}}} - \frac{\xi_{\bf p}}{E_{\bf p}}  \right ) \right ]
\frac{n_{{\bf p}+{\bf q}} - n_{\bf p}}{ \omega_{n}^{2} + (E_{{\bf p}+{\bf q}} - E_{\bf p})^{2}}
\nonumber 
\\
& \quad
-
\sum\limits_{\bf p} 
\frac{1}{2} \left [ (E_{{\bf p}+{\bf q}} + E_{\bf p}) \left ( 1 + \frac{\xi_{{\bf p}+{\bf q}}\xi_{\bf p}}{E_{{\bf p}+{\bf q}}E_{\bf p}} \right ) 
+ i \omega_{n} \left ( \frac{\xi_{{\bf p}+{\bf q}}}{E_{{\bf p}+{\bf q}}} + \frac{\xi_{\bf p}}{E_{\bf p}}  \right ) \right ]
\frac{1 + n_{{\bf p}+{\bf q}} + n_{\bf p}}{ \omega_{n}^{2} + (E_{{\bf p}+{\bf q}} + E_{\bf p})^{2}} , 
\label{gapless Hartree-Fock-BogoliubovPi11fp}
\\
\Pi_{12} (q) = & 
- \sum\limits_{\bf p} 
\frac{1}{2} \Delta \left [ \frac{\xi_{{\bf p}+{\bf q}}}{E_{{\bf p}+{\bf q}}E_{\bf p}} (E_{{\bf p}+{\bf q}} - E_{\bf p}) + \frac{ i \omega_{n} }{E_{\bf p}} \right ]
\frac{n_{{\bf p}+{\bf q}} - n_{\bf p}}{\omega_{n}^{2} + (E_{{\bf p}+{\bf q}} - E_{\bf p})^{2}}
\nonumber 
\\
& \quad
+ \sum\limits_{\bf p} 
\frac{1}{2} \Delta \left [ \frac{\xi_{{\bf p}+{\bf q}}}{E_{{\bf p}+{\bf q}}E_{\bf p}} (E_{{\bf p}+{\bf q}} + E_{\bf p}) + \frac{ i \omega_{n} }{E_{\bf p}} \right ]
\frac{1 + n_{{\bf p}+{\bf q}} + n_{\bf p}}{ \omega_{n}^{2} + (E_{{\bf p}+{\bf q}} + E_{\bf p})^{2}} , 
\label{gapless Hartree-Fock-BogoliubovPi12fp}
\\
\Pi_{14} (q) = & 
\sum\limits_{\bf p} 
\frac{1}{2} \frac{\Delta^{2}}{E_{{\bf p}+{\bf q}}E_{\bf p}}
\left [ 
(E_{{\bf p}+{\bf q}} - E_{\bf p})
\frac{n_{{\bf p}+{\bf q}} - n_{\bf p}}{ \omega_{n}^{2} + (E_{{\bf p}+{\bf q}} - E_{\bf p})^{2}}
- (E_{{\bf p}+{\bf q}} + E_{\bf p})
\frac{1 + n_{{\bf p}+{\bf q}} + n_{\bf p}}{ \omega_{n}^{2} + (E_{{\bf p}+{\bf q}} + E_{\bf p})^{2}}
\right ] , 
\label{gapless Hartree-Fock-BogoliubovPi14fp}
\\
\Pi_{22} (q) = & 
\sum\limits_{\bf p} 
\frac{1}{2} \left [ (E_{{\bf p}+{\bf q}} - E_{\bf p}) \left ( 1 + \frac{\xi_{{\bf p}+{\bf q}}\xi_{\bf p}}{E_{{\bf p}+{\bf q}}E_{\bf p}} \right ) 
+ i \omega_{n} \left ( \frac{\xi_{{\bf p}+{\bf q}}}{E_{{\bf p}+{\bf q}}} + \frac{\xi_{\bf p}}{E_{\bf p}}  \right ) \right ]
\frac{n_{{\bf p}+{\bf q}} - n_{\bf p}}{ \omega_{n}^{2} + (E_{{\bf p}+{\bf q}} - E_{\bf p})^{2}}
\nonumber 
\\
& \quad
+
\sum\limits_{\bf p} 
\frac{1}{2} \left [ (E_{{\bf p}+{\bf q}} + E_{\bf p}) \left ( 1 - \frac{\xi_{{\bf p}+{\bf q}}\xi_{\bf p}}{E_{{\bf p}+{\bf q}}E_{\bf p}} \right ) 
+ i \omega_{n} \left ( \frac{\xi_{{\bf p}+{\bf q}}}{E_{{\bf p}+{\bf q}}} - \frac{\xi_{\bf p}}{E_{\bf p}}  \right ) \right ]
\frac{1 + n_{{\bf p}+{\bf q}} + n_{\bf p}}{ \omega_{n}^{2} + (E_{{\bf p}+{\bf q}} + E_{\bf p})^{2}} , 
\label{gapless Hartree-Fock-BogoliubovPi22fp} 
\end{align}
\end{widetext}
where $\xi_{\bf p} \equiv \varepsilon_{\bf p} + \Delta$, $\Delta \equiv U n_{0}$, $E_{\bf p} \equiv \sqrt{\varepsilon_{\bf p} (\varepsilon_{\bf p} + 2 \Delta )}$, 
and $n_{\bf p} \equiv 1/ [ \exp{( E_{\bf p} /T)} - 1]$.  
\par
At $T \geq T_{\rm c}$, the polarization functions are given by 
\begin{align}
\Pi_{11} (q) = & 
- \sum\limits_{{\bf p}}
\frac{1 + 
n_{{\bf p} + {\bf q}}' 
+ 
n_{{\bf p} }' 
}
{
\varepsilon_{{\bf p} + {\bf q} } + \varepsilon_{{\bf p} }  +2 \Sigma_{11} (0)  
- 2 \mu - i \omega_{n}
}, 
\\
\Pi_{22} (q) = & 
 \sum\limits_{{\bf p}}
\frac{
n_{{\bf p} + {\bf q}}'
-
n_{{\bf p}}'
}
{
\varepsilon_{{\bf p} + {\bf q}} - \varepsilon_{{\bf p}} - i \omega_{n}
} , 
\end{align}
where $n_{\bf p}' = 1/ ( \exp{ \{ [ \varepsilon_{\bf p} + \Sigma_{11} (0) - \mu ] / T  \}  }- 1 )$.

\bibliography{BosePaper.bib}

\end{document}